 \documentclass[preprint]{aastex}

\shortauthors{Moon, Eikenberry, \& Wasserman}
\shorttitle{X-RAY FLARES from LMC X-4}

\begin{document}

\title{The Evolution Of LMC X-4 Flares:  \\
Evidence For Super-Eddington Radiation Oozing \\
Through Inhomogeneous Polar Cap Accretion Flows ?}

\author{Dae-Sik Moon, Stephen S. Eikenberry\altaffilmark{1}, \& Ira M. Wasserman}
\affil{Department of Astronomy, Cornell University, Ithaca, NY 14853; \\
moon@astro.cornell.edu, eiken@astro.cornell.edu, ira@astro.cornell.edu}

\altaffiltext{1}{also Department of Astronomy, University of Florida, Gainesville, FL 32611}

\begin{abstract}
We present the results of two extensive {\it Rossi X-ray Timing Explorer} 
observations of large X-ray flaring episodes from the high-mass X-ray binary 
pulsar LMC X-4.
Light curves during the flaring episodes comprise bright peaks embedded 
in relatively fainter regions, 
with complex patterns of recurrence and clustering of flares.
We identify precursors preceding the flaring activity.
Pulse profiles during the flares appear to be simple sinusoids,
and pulsed fractions are proportional to the flare intensities.
We fit Gaussian functions to flare peaks to estimate the mean
full-width-half-maximum to be $\sim$68 s.
Significant rapid aperiodic variability exists up to a few hertz during the flares,
which is related to the appearance of narrow, spiky peaks in the light curves.
While spectral fits and softness ratios show overall
spectral softening as the flare intensity increases,
the narrow, spiky peaks do not follow this trend.
The mean fluence of the flare peaks is (3.1 $\pm$ 2.9) $\times$ 10$^{40}$ ergs
in the 2.5--25 keV energy range, with its maximum at $\sim$1.9 $\times$ 10$^{41}$ ergs.
The flare peak luminosity reaches up to (2.1 $\pm$ 0.2) $\times$ 10$^{39}$ ergs s$^{-1}$, 
far above the Eddington luminosity of a neutron star.
We discuss possible origins of the flares,
and we also propose that inhomogeneous accretion columns onto the neutron star
polar caps are responsible for the observed properties.
\end{abstract}

\keywords{accretion, accretion disks --- pulsars: individual (LMC X-4) --- stars: neutron --- X-rays: bursts --- X-rays: stars}

\section{INTRODUCTION}

Out of $\sim$200 known neutron star/black hole X-ray binaries,
more than 40 sources have been observed with a wide range of 
coherent spin period of $\sim$0.07--1400 s, 
identifying the central source as a rotating, highly magnetized
neutron star powered by accretion (van Paradijs 1995; Bildsten et al. 1997).
These accretion-powered X-ray binary pulsars (hereafter, X-ray pulsars)
often show abrupt increases in X-ray luminosity known as ``bursts" or ``flares."
About 30 \% of the X-ray pulsars have been observed to show bursts and/or flares,
including (1) the low-mass X-ray binary 4U 1626--67 (McClintock et al. 1980),
(2) the bursting pulsar GRO J1744--28 (Fishman et al. 1995; Kouveliotou et al. 1996),
(3) the high-mass X-ray binaries such as LMC X-4 
(e.g., Levine et al. 1991; Woo, Clark, \& Levine 1995;
Levine, Rappaport, \& Zojcheski 2000; Moon \& Eikenberry 2001)
and SMC X-1 (Angelini, Stella, \& White 1991; Moon, Eikenberry, \& Wasserman 2002),
and (4) the transient Be binary system EXO 2030+375 (Parmar et al. 1989).
Although they are easily identified 
as sudden increases of X-ray luminosity in most cases, 
the detailed properties of the bursts/flares differ much
from source to source.
For instance, whereas bursts from the high-mass system SMC X-1 have been detected
only once despite extensive studies of the source 
\footnote{SMC X-1 has been identified as a flaring X-ray pulsar very recently (Moon et al. 2002).}, 
another high-mass system LMC X-4 has been observed with large, semi-regular flaring episodes
over the last two decades by various X-ray satellites.
This diversity in bursts/flares suggests that many different
mechanisms are responsible for the observed phenomena.

Most theoretical studies of the origin of bursts/flares 
from X-ray pulsars have relied upon one of three following mechanisms:
(1) accretion disk instabilities, 
(2) nuclear burning on the surface of the neutron star, and
(3) non-uniform stellar winds from mass-donating companions.
Obviously the most well known case for the accretion disk instability 
is the X-ray bursts from the bursting pulsar GRO 1744--28,
for which a thermal-viscous instability in the accretion disk has been 
investigated to explain the origin of the bursts (e.g., Cannizzo 1996, 1997).
The X-ray bursts from GRO 1744--28 share many properties with 
the Type II X-ray bursts from the Rapid Burster (Lewin et al. 1996)
and the X-ray flares from SMC X-1 (Moon et al. 2002).
On the other hand,
the presence of a He-rich dwarf companion led Brown \& Bildsten (1998) to suggest the possibility that 
$\sim$1000-s flares from 4U 1626--67 are caused by carbon burning under the neutron star surface.
Taam, Fryxell, \& Brown (1988) attributed the origin of the
flares from EXO 2030+375 to an accretion disk instability 
caused by the inhomogeneity in the high-velocity ($\sim$550 km s$^{-1}$) winds 
from a Be star companion.

LMC X-4 is a persistent, disk-fed, high-mass system 
with a pulsational and an orbital period of $\sim$13.5 s and $\sim$1.4 d,
respectively (Kelley et al. 1983).
It is one of a few X-ray binaries showing a super-orbital period
($\sim$30.3 day; Ilovaisky et al. 1984) possibly caused by 
a precessing, tilted accretion disk.
The optical companion is a 14th magnitude 
O-type star of $\sim$15 $M_{\odot}$ (Sanduleak \& Philip 1977).
Over the last two decades, LMC X-4 has exhibited
large X-ray flaring episodes which have made it one of the most
regular and representative flaring X-ray sources.
During the flares, the X-ray spectrum softens considerably,
and the pulse profiles become simple, sinusoidal (Levine et al. 1991, 2000).
Recently, Moon \& Eikenberry (2001)
reported two relatively long-time scale structures in the flares:
quasi-periodic oscillations of $\sim$0.65--1.35 and
$\sim$2--20 mHz. 
They also found that the amplitudes of the flares
are almost exactly modulated by those of the 
neutron star's coherent pulsations (or vice versa), 
indicating that the flares occur near the magnetic pole of the neutron star.
Although the LMC X-4 flares have been observed for many years
with various X-ray satellites, no convincing mechanism has been proposed to 
explain the origin of the flares (e.g., Levine et al. 2000).

On the other hand, LMC X-4 flares have been known
as super-Eddington phenomena:
sometimes the flare X-ray luminosity reaches up to $\sim$10$^{39}$ ergs s$^{-1}$
(while its normal state luminosity is $\sim$2 $\times$ 10$^{38}$ ergs s$^{-1}$,
comparable to the Eddington luminosity of the
$\sim$1.4 $M_{\odot}$ neutron star of LMC X-4).
Because the key physical parameters of LMC X-4 --
such as mass, distance, and magnetic field  --
are relatively well determined to be 1.4 $\pm$ 0.3 $M_{\odot}$ (Levine et al. 1991),
50 $\pm$ 10 kpc (e.g., Kov\'{a}cs 2000), and
$\sim$10$^{13}$ G (La Barbera et al. 2001), respectively,
the LMC X-4 flares may provide a rare example for studying
super-Eddington phenomena thoroughly without assuming the key parameters.
This is not the case for most super-Eddington sources
owing to relatively large uncertainties in their distances and masses.
Considering the recent important controversy over the existence 
of the so-called ``intermediate-mass black holes'' (e.g., King et al. 2001; Begelman 2001, 2002),
for which mass ($\sim$10$^2$--10$^4$ $M_{\rm \odot}$) is determined under the assumption 
that the observed X-ray luminosity is the Eddington luminosity, 
the investigation of the origin of the super-Eddington radiation
from the LMC X-4 flares can also be useful to investigate the origin of the claimed 
intermediate-mass black holes.

In this paper, we present by far the most extensive observations of the LMC X-4 flares
made with the {\it Rossi X-ray Timing Explorer} (RXTE). 
We describe the RXTE observations and our data analyses in \S 2;
we present the results of timing and spectral analyses in \S 3.
We discuss the possible origin of the LMC X-4 flares, 
as well as their super-Eddington radiation, in \S 4,
We summarize our results in \S 5.

\section{Observations and Data Analyses}

RXTE (Bradt, Rothschild, \& Swank 1993) observed two large flaring episodes of LMC X-4
on 1996 August 19 and 1999 December 19 during its long-exposure (49 and 42 h) observations of the source.
The HEAsoft\footnote{http://heasarc.gsfc.nasa.gov/docs/software/lheasoft/}
package (version 5.1) was used for analyzing the data from the
Proportional Counter Array (PCA; Jahoda et al. 1996).
The data within 30 minutes after passages
through the South Atlantic Anomaly and/or 
with a higher ($>$ 0.1) electron ratio were ignored.
Only the data obtained with three or more active Proportional Counter Units (PCUs) 
were used for timing analyses; only the data obtained with all five PCUs were used for spectral analyses.
The photon arrival times from the Good Xenon mode were transformed to the solar system barycenter 
using the JPL DE400 ephemeris.
The Very Large Event Models of Epoch 3 and 4 were used to subtract backgrounds
for the data obtained in 1996 and 1999, respectively.
The Standard 2 data obtained only from the top xenon layers of 
the PCUs 0, 1, 2, and 4 were used in spectral fits due to responsivity problems of PCU 3 
(R. Remillard 2002, private communication).
Only the spectrum between 2.5--25 keV range was considered for the same reason, and 
a systematic uncertainty of 1 \% was assumed in spectral fits.
A total of 45 data segments (24 from the 1996 observations and 21 from the 1999 ones) of $\sim$1-h length was obtained, 
and flares were observed in 4 and 6 segments of the two observations.

\section{Results}

\subsection{Light curves and Rapid Aperiodic Variability}

\subsubsection{Light curves}

Because no reliable ephemeris of the $\sim$30.5-d super-orbital
period of LMC X-4 around the two observations reported here was available,
we used the RXTE All Sky Monitor (ASM; Bradt et al. 1993) data 
to examine the phase of the super-orbital motion.
Figure 1 shows the ASM light curves around the two observations, 
indicating that they are likely for the high state of the source.

Figure 2 presents background-subtracted light curves of the PCA data
-- (a) for the 1996 observations; (b) for the 1999 ones.
The Modified Julian Dates (MJDs) of the start of the two 
light curves are 50314.07905 and 51531.48489, respectively.
The binary orbital phase in the upper x-axis was
determined by the LMC X-4 ephemeris of Levine et al. (2000).
The flares are easily identified in the first four data segments 
of Figure 2a and six data segments of Figure 2b as significant 
increases in the photon count rates.
Whereas the first half of the data of the 1996 observations
shows strong flaring activity in the orbital phases $\phi$ $\simeq$ 0.26--0.6,
the second half shows no noticeable activity at the same orbital phases
(although some phases are missing due to data gaps).
The LMC X-4 eclipse appears around $\phi$ = 1.0
as an apparent decrease in the photon count rates.
The 1999 observations show flaring activity at $\phi$ $\simeq$ 0.4--0.8.  
The mean photon count rates of the normal state is
$\sim$44 and $\sim$54 counts/sec/PCU for the 1996 and the 1999 observations, respectively,
and that of the eclipse is $\sim$13 counts/sec/PCU.
(We refer to the normal state in this paper as a state absent of 
flaring activity and eclipse.)

Figure 3 presents the magnified light curves of the ten data segments containing flares with 4-s resolution.
We shall call them ``FL1" to ``FL10" in time order.
Figure 3 shows the complex flaring activities of LMC X-4 summarized as follows.
First, the light curves with strong flares (i.e., FL1--5)
appear to consist of multiple peaks embedded in the relatively 
low-intensity regions that are still brighter than the normal state;
those with weak flares (i.e., FL6--10) show 
simple, small peaks connected through the regions of which
brightness is comparable to that of the normal state.
Secondly, FL1 and FL2 clearly show recurrences of intense flares;
FL3 and FL5 show a strong concentration of intense flares with a time scale of $\sim$2000 s.
Finally, while most of the flare peaks show symmetric, Gaussian-like profiles, 
the largest ones (e.g., those at $t$ $\simeq$ 3240 s of FL3,
$t$ $\simeq$ 1770 s of FL4, and $t$ $\simeq$ 2010 s of FL5)
appear to show rather asymmetric profiles with a slow decay.

For quantitative analyses, 
we performed multi-component Gaussian fits to the flare peaks in Figure 3.
Table 1 summarizes parameters of the 68 Gaussian-fitted flare peaks
of which peak intensity is greater than 90 counts/sec/PCU.
The mean peak intensity and full-width-half-maximum (FWHM) 
are 172 $\pm$ 103 counts/sec/PCU and
68 $\pm$ 31 s, respectively.
The median and mean values of the reduced chi-square (= $\chi^2_{\nu}$) 
of the fits are 1.6 and 3.8 $\pm$ 5.7, respectively.
This indicates that while most of the flare peaks are fitted reasonably well
(i.e., $\chi^2_{\nu}$ $<$ 2) with a Gaussian function, 
some significantly deviate from a Gaussian.
If we exclude the six flare peaks with $\chi^2_{\nu}$ greater than 10
(including the three aforementioned largest flare peaks that apparently show slow decays), 
the mean $\chi^2_{\nu}$ decreases to 2.2 $\pm$ 1.9.  
The mean fluence of the flare peaks is (3.1 $\pm$ 2.9) $\times$ 10$^{40}$ ergs,
with the maximum at 1.9 $\times$ 10$^{41}$ ergs.
The fluence is the integrated luminosity over the FWHM, and the luminosity was
computed by spectral fits (see \S3.2).
(All the errors quoted here represent 1-$\sigma$ rms deviation.)

In Figure 3, the light curves FL2, FL3, and FL4 begin with a normal state,
while other light curves begin with a flaring state.
The mean photon count rates of the first 500 s of FL2, FL3, and FL4 are 
$\sim$35, $\sim$39, and $\sim$40 counts/sec/PCU, respectively, 
similar to those of the normal state.
This offers an opportunity to examine the light curve transition
from a normal states to a flaring state. 
In fact, the magnified views around the beginnings of the flares in FL2, FL3, and FL4 (Figure 4)
identify the existence of the rather flat precursors of the flares 
lasting for a few minutes just before abrupt increases in photon count rates.

In contrast with the complicated differences in the flare light curves (Figure 3),
the pulse profiles of FL1--10 show a remarkable similarity (Figure 5),
all having simple, sinusoidal profiles.
Figure 6 compares the pulsed fraction with the peak intensity of the 10 pulse profiles in Figure 5.
The pulsed fraction is proportional to the peak intensity
of the pulse profile with a linear correlation coefficient of $\sim$0.76.

\subsubsection{Rapid Aperiodic Variability}

In order to examine the possible existence of any rapid variability associated with the flares, 
we present 45-s light curves of FL1--10 with 0.0625-s resolution around 
the brightest peak in each light curve (Figure 7).
The light curves are modulated by $\sim$13.5-s pulsations of the neutron star of LMC X-4, 
and exhibit strong narrow peaks of which intensities are significantly greater than those
expected from the Poisson noise of the given light curve.
The Leahy-normalized power-density spectra (PDSs) of the light curves 
of the central 10-s segments reveal the existence of significant powers up to a few hertz,
especially for the bright flares such as FL2--5.
(Figure 8; We only used the central 10-s segments in order to avoid 
the contamination by the source's pulsations in PDSs.
Note that the Leahy power owing to a Poisson noise is 2 [Leahy et al. 1983]).)
We investigated this possible correlation between the existence of
the significant powers in the PDSs and the flare intensity more as follows.
First, we divided the FL1--10 light curves (Figure 3) into segments of which
lengths are equal to the pulsational period ($\sim$13.5 s) of LMC X-4
determined by Fourier transformations of the light curves.
Next, we computed the Leahy-normalized PDSs of the all $\sim$13.5-s segments,
as well as their integrated intensities.
The correlation between the flare intensity and the existence of
the significant powers in PDSs is evident in Figure 9, 
where we compare the distributions of the integrated intensities 
with those of the significant powers ($>$ 10) (in PDSs) of the $\sim$13.5-s segments.
This confirms that the rapid aperiodic variability during the flares 
increases along with the flare intensity, which sometimes appears
as strong narrow peaks in the light curves (Figure 7).

\subsection{Spectral Evolution of the Flares}

\subsubsection{Softness Ratio Distribution}

We investigated the spectral evolution of the flares via the softness ratio,
which we defined to be the ratio of the soft X-ray (2--8 keV) 
photon count rates to those of the hard X-ray (10--20 keV). 
Figure 10 compares the total photon count rates (integrated over the 2--25 keV range)
of the flares with the softness ratios obtained with 32-s resolution, 
identifying a strong linear correlation between them with 
a linear correlation coefficient of $\sim$0.95.

However, Figure 11a, which is the same as Figure 10 but obtained with 0.0625-s resolution,
shows that the strong correlation identified with 32-s resolution
disappears when time resolution improves -- the linear correlation coefficient decreases to $\sim$0.30.
The data points with large photon count rates (so with small Poisson noise)
show more significant deviations from the linear correlation 
(identified  with 32-s resolution) than those with small photon count rates,
indicating that the deviation is not simply owing to the increased Poisson noise 
caused by the improved time resolution.
We investigated this further via comparison of the 
correlations obtained from two separate groups of the data:
one is the ``faint group" having small total photon count rates; 
the other is the ``bright group" having large total photon count rates.
We chose the faint group as the group of data points of which total photon count rates belong to the lower 90 \%; 
the bright group of which total photon count rates belong to the upper 1 \%.
The dotted and dashed lines in Figure 11a represent the correlations obtained 
in the faint and bright groups, respectively.
The correlations obtained in the two groups are quite distinctive:
while the faint group (dotted line) shows a positive correlation with a slope 
similar to that obtained with 32-s resolution (Figure 10), 
the bright group (dashed line) does not show any apparent correlation,
indicating that the strong linear correlation between the flare intensities and the 
softness ratios obtained with 32-s resolution does not hold for 
this component (i.e., the narrow, spiky peaks in Figure 7).

In order to examine the effect of the increased Poisson noise 
caused by the improved time resolution in Figure 11a more thoroughly,
we used a Monte Carlo simulation by re-sampling the each data point 
obtained with 32-s resolution (Figure 10) with 0.0625-s resolution.
We re-sampled the 32-s resolution data points using a Gaussian distribution 
with the 1$\sigma$ standard deviation equal to the Poisson noise of 0.0625-s resolution.
We performed the simulation 1000 times, and present a typical result in Figure 11b.
Although the correlation coefficients of the three lines representing
the all data points (solid line, $\sim$0.20), 
the faint group (dotted line, $\sim$0.09), and the bright group (dashed line, $\sim$0.25)
are much smaller than that obtained with 32-s resolution ($\sim$0.95),
they show very similar slopes to that obtained with 32-s resolution and
no apparent difference among them.
This contrasts with the real data (Figure 11a), 
demonstrating that the simulated data are unable to produce the decreases 
in the softness ratios of the bright group of the real data
(while the correlation coefficients of the simulated data decrease 
due to the increased Poisson noise caused by the improved time resolution, 0.0625 s).
Therefore, the absence of the linear correlation between the softness ratios 
and the total photon count rates in the bright group of the real data seems to be of statistical significance.
Table 2 summarizes the correlation coefficients and slopes of the lines in Figures 10 and 11.

The systematic difference in the softness ratios of the bright and faint groups of
the real data is more clearly identified in Figure 12, 
where we compare the binned softness ratios of the real data (dotted line) in Figure 11a with
those of the simulated data (solid line) in Figure 11b.
For this, we made 100 bins of the total photon count rates 
(each bin with $\sim$1230 data points), 
and computed the mean softness ratios in each bin.
As we see in Figure 12, the softness ratios of the real data with small total 
photon count rates ($<$ 300 counts/sec/PCU) are almost identical to those of the simulated data;
however, the real data show a clear systematic decrease in the softness ratios
as the total photon count rates increase ($>$ 300 counts/sec/PCU).
About 85 \% of the data have total photon count rates less than 300 counts/sec/PCU.

\subsubsection{Spectral Fits}

In order to study the spectral evolution of the flares more thoroughly, 
we obtained flare spectra over 32-s interval, 
and fitted them to a model spectrum consisting of a power law, a black body, and a Gaussian component, 
as well as a component for photoelectric absorption due to the intervening interstellar matter.
The power law and Gaussian components represent 
non-thermal magnetospheric emission and iron line emission, respectively, 
which is typical of X-ray pulsar (e.g., White, Swank, \& Holt 1983).
The black body component represents the thermal emission associated with
the neutron star polar cap and the accretion columns during the flares.
The model spectrum is given by
\begin{equation}
I(E) ={ {\rm exp}[- \sigma (E) N_{\rm H}] 
       \times \{ I_{\rm P} E^{- \alpha} + BB(E) + G(E) \} } \\
      {\rm \; \; \; photons \; cm^{-2} \; s^{-1} \; keV^{-1}} 
\end{equation}
where $E$ is the photon energy in keV;
$\sigma$(E) and $N_{\rm H}$ are the 
photoelectric cross section and the hydrogen nuclei column density of the intervening matter;
$I_{\rm P}$ and  $\alpha$ are the norm and the power-law index of the power-law component.
$BB(E)$ is the black body function of temperature $kT$;
$G(E)$ is the Gaussian function for the iron line emission with its center at $E_{\rm G}$.
The lower limit of $N_{\rm H}$ was fixed to be 
3 $\times$ 10$^{20}$ cm$^{-2}$ based on the previous results 
(Woo et al. 1995, and references therein).
We obtained the mean $\chi^2_{\nu}$ of 0.99 $\pm$ 0.22 with $\sim$43 degrees of freedom
from spectral fits between 2.5 and 25 keV.

Figure 13 presents the distributions of the resulting power index ($\alpha$, 13a)
and the black body temperature ($kT$, 13b) along with the flux 
obtained from the model fits in the 2.5--25 keV range.
The power index increases as the flux increases;
the black body temperature decreases on the contrary.
The possible saturation in the decrease of the black body temperature 
may be due to the lower energy limit of the fits (= 2.5 keV).
Figure 13 shows that both the power-law and black body components
soften as the flare intensity increases,
which is consistent with the softness ratio distribution obtained with 32-s resolution (Figure 10).
(We tried many different model spectra,
e.g., adding a high-energy cutoff to the power-law component,
subtracting the Gaussian component or the black body component, 
and fixing the hydrogen nuclei column density.
The fits to these trial spectra gave us a moderately reasonable $\chi^2_{\nu}$ ($<$ 2).
The trends identified in Figure 13 always existed in the spectral fits to these trial model spectra.
The fits to the model spectrum of Eqn. [1] yielded the least $\chi^2_{\nu}$.)
Table 3 lists the luminosities of the precursors (Figure 4) 
and the brightest flare peak of each FL1--10 (Figure 3), 
as well as the fluence of FL1--10. 
We used a distance of 50 kpc to LMC X-4 to estimate the luminosity.
Figure 14 presents a comparison between the best-fit model spectra (solid lines) 
and the observed ones (crosses) of the brightest flare peak of each FL1--10. 
Table 4 summarizes the best-fit parameters of Figure 14.

We compared the spectrum from the normal state with that 
of the brightest flare peak in all FL1--10 
(i.e., the peak at $t$ $\simeq$ 1770 s of FL4 in Figure 3).
We used three different model spectra:
(1) the model spectrum given by the Eqn. (1), 
(2) a model spectrum consisting only of a power-law component and a Gaussian component,
(3) the same model spectrum of (2) but with a high-energy cutoff for the power-law component.
For the case (3), the model spectrum is given by
\begin{equation}
I(E) ={ {\rm exp}[- \sigma (E) N_{\rm H}]
       \times \{ I_{\rm P} E^{- \alpha} f_{hi}(E) +  G(E) \} } \\
      {\rm \; \; \; photons \; cm^{-2} \; s^{-1} \; keV^{-1}}
\end{equation}
where
$$ f_{hi}(E) = \cases{1,&                            $E < E_c$              \cr
                      exp(- {(E - E_c) \over E_f}),& $E \geq E_c$,  \quad  \cr}
$$
represents the high-energy cutoff for the power-law component.
Figure 15 compares the three best-fit model spectra (solid lines) 
with the observed ones (crosses),
and Table 5 summarizes the best-fit parameters.
All the spectral fits to the three model spectra gave us acceptable
$\chi^2_{\nu}$ ($<$ 2). 
We noted that the fitted values for the high-energy cutoff component
of the normal state were very close to those previously obtained
(16.1 $\pm$ 0.4 and 35.6 $\pm$ 6.7 keV; Woo et al. 1996),
whereas those for the flare peak were considerably different.
In order to examine the significance of this change in the high-energy cutoff component, 
we fit the flare peak spectrum to the model spectrum given by the Eqn. (2)
with the high-energy cutoff component fixed to the values obtained 
from the normal state: $E_{\rm c}$ = 16.1 keV and $E_{\rm f}$ = 27.4 keV.
The resulting best-fit parameters are: $N_{\rm H}$ = (6.5 $\pm$ 0.7) $\times$ 10$^{22}$ cm$^{-2}$,
$\alpha$ = 3.3 $\pm$ 0.2, and $E_{\rm G}$ = 5.8$^{-0.6}_{+0.4}$ keV (with $\chi^2_{\nu}$ $\simeq$ 1.3),
similar to the values in Table 5.
Therefore, the changes in the high-energy cutoff component between the normal state and 
flares do not seem to be significant.
The mean luminosity of the normal state luminosity is $\sim$1.5--2.0 $\times$ 10$^{38}$ ergs s$^{-1}$.


\section{Discussion}

\subsection{Possible Origin of the Flares}

\subsubsection{Thermonuclear Burning}

Levine et al. (2000) discussed the possibility of 
carbon burning under the neutron star surface as the origin of the LMC X-4 flares.
According to their discussion, the nuclear burning scenario can roughly explain 
several features of the flares, including the luminosity, the daily recurrence time, 
the rapid rise and gradual decay, and the spectral softening.
Shortcomings are an absence of any mechanism which can explain 
the $\sim$100--200 s substructure and the lack of enough carbon material 
needed for powering the flares when the accreting material is 
mostly H-rich material (Brown \& Bildsten 1998).
Although it might be possible to attribute the $\sim$100--200 s substructure 
to an accretion disk origin (Moon \& Eikenberry 2001),
the results of this paper are not supportive of the nuclear burning mechanism 
as the origin of the flares.
First, the precursors found in this paper (Figure 4) do not reconcile readily with the nuclear burning scenario,
because the higher accretion rate during the precursor on the polar cap region would suppress the
thermonuclear instability (e.g., Bildsten \& Brown 1997; Brown \& Bildsten 1998).
We note that Type I X-ray bursts, 
which are known to be caused by nuclear flashes on the neutron star surface
in low-mass X-ray binaries,
have occasionally been observed with precursors of a few seconds (e.g., Lewin, Vacca, \& Basinska 1984).
However, the burst precursors are distinctive from the precursors of the LMC X-4 flares of this paper.
The burst precursor is generally followed by a normal state of a few seconds
due to a rapid expansion and subsequent cooling of the neutron star photosphere,
which is clearly different from the LMC X-4 flares.
In addition, while many Type I X-ray bursts have been observed with spectral 
softening during the decay,
the LMC X-4 flares show the opposite phenomena: the spectrum is hardening during the  decay.
The very strong magnetic field strength of LMC X-4 ($\sim$10$^{13}$ G; La Barbera et al. 2001)
also makes the nuclear burning scenario less likely --
no X-ray pulsars have been observed with Type I X-ray bursts
except for rare accreting millisecond pulsars (e.g., SAX J1808.4--3658; in 't Zand et al. 1998).
Finally, the observed spectra of the flares are far from thermal,
although the thermal spectra from a highly magnetized object
may be very different from those of non-magnetized objects.
Therefore, it is very difficult to explain the LMC X-4 flares 
in terms of the nuclear burning scenario.

\subsubsection{Viscous Instability}

We note that X-ray bursts/flares from the X-ray pulsars GRO J1744--28 and SMC X-1 
have been explained as results of viscous instabilities in the accretion disk
(Cannizzo 1996, 1997; Li \& van den Heuvel 1997; Moon et al. 2002).
If the radiation pressure is comparable to the gas pressure around the inner-disk radius,
the accretion disk is susceptible to a viscous instability.
Under the classical $\alpha$-disk configuration (Shakura \& Sunyaev 1973),
this requires that the transition layer between the radiation-pressure-dominated and the gas-pressure-dominated
regions be comparable to the inner-disk radius.
In order to satisfy this condition, 
the magnetic moment of the neutron star is expected to be (Shapiro \& Teukolsky 1983)
\begin{equation}
\mu_{30} \simeq 4 \times 10^{-3} \; \dot M_{17}^{11/6}
\end{equation}
where $\mu_{30}$ is the magnetic moment of a neutron star in unit of 10$^{30}$ G cm$^3$,
and $\dot M_{17}$ is the mass accretion rate in unit of 10$^{17}$ g s$^{-1}$.
We assumed 1.4 $M_{\odot}$ and 10 km for the mass and radius of the neutron star, respectively,
and obtained $\mu_{30}$ $\simeq$ 0.3 with $\dot M_{17}$ $\simeq$ 10.
This is significantly smaller than the value estimated from observations
($\mu_{30}$ $\simeq$ 11; La Barbera et al. 2001),
indicating that the viscous instability is difficult to grow in the accretion disk 
around LMC X-4 with the $\alpha$-disk configuration. 
This also confirms that it is unlikely that radiation-pressure-dominated regions 
can exist around highly-magnetized neutron stars (i.e., X-ray pulsars).

\subsubsection{Plasma Rayleigh-Taylor Instability and Stellar Winds}

LMC X-4 is known to be in a spin equilibrium state (Naranan et al. 1985; Woo et al. 1995; La Barbera 2001) 
in which the magnetospheric radius is comparable to the corotation radius of the source.
In this case, it has been suggested that the plasma Rayleigh-Taylor instability
(which is known as the Kruskal-Schwarzschild instability)
can develop to result in X-ray bursts from the 
neutron star X-ray binaries (Baan 1977, 1979).  According to this scenario,
accreting matter can be accumulated outside the magnetosphere when the neutron star spins faster
than the inner accretion disk owing to the centrifugal barrier of the magnetosphere.
Once the density of the accumulating matter exceeds the critical value, 
the effective gravity becomes larger, driving the magnetosphere inside the corotation radius,
and an abrupt increase in the accretion onto the neutron star can occur.
Based on this mechanism, Apparao (1991) explained the origin of the six flares observed from EXO 2030+375.
We note that the flares of EXO 2030+375 share similar properties with the LMC X-4 flares, 
including the $\sim$3--4 h flare recurrence time, the $\sim$1--2 h flare duration, 
the high fluences of the flares ($>$ 10$^{40}$ ergs), the very similar precursors,
the rapid rise with gradual decay, and the absence of any significant spectral change
(Parmar et al. 1989; Parmar, White, \& Stella 1989).
Similar to LMC X-4, $\sim$10$^{13}$ G magnetic field strength
is expected from EXO 2030+375 (Apparao 1991).
These similar properties are suggestive that the same mechanism may be responsible for 
the flares from the both objects.

Finally, we investigate the possibility that the LMC X-4 flares  
are caused by non-uniform stellar winds from the O-type companion.
One noticeable result is that, 
based on the estimation of stellar mass loss rate, 
Boroson et al. (1999) found that the gravitational capture of stellar winds could power 
$\sim$3 $\times$ 10$^{37}$ ergs s$^{-1}$ luminosity of LMC X-4,
which is comparable to the excess luminosities of the flare precursors
over the normal states (Table 3).
In fact, several early type stars have been observed 
with non-uniform stellar winds of $\sim$2--3 h periods (Vogt \& Penrod 1983; Kudritzki \& Puls 2000),
which is similar to the time interval of the 
three consecutive flares of the LMC X-4 (FL1--3; Figure 2 and 3).
(The flares FL6--10 are also consecutive, 
but it is difficult to estimate flare recurrence time from these flares.)
Additionally, the roughly estimated recurrence period of the flaring episode of LMC X-4
is compatible to its orbital period, $\sim$1.4 d.
Based on these, one can imagine a very speculative scenario in which the LMC X-4 flare precursors  
are caused by the capture of material in the non-uniform stellar winds from the companion star 
related to binary orbital motion, 
and then the precursor induces instabilities in the accretion disk
(such as the plasma Rayleigh-Taylor instability) by satisfying the conditions for the instability to develop.

\subsection{Softness Ratio Distributions and Inhomogeneous Polar Cap During Super-Eddington Flares}

\subsubsection{Softness Ratio Distributions and Inhomogeneous Polar Cap Accretion Flows}

As identified by the analysis with 0.0625-s resolution, 
the LMC X-4 flares appear to consist of two distinctive components.
The first one, which is the major component corresponding to $\sim$85 \% of the total data,
has relatively small total photon count rates ($<$ 300 counts/sec/PCU); 
the second one, which is the minor component corresponding to the remaining $\sim$15 \%,
has large total photon count rates ($>$ 300 counts/sec/PCU).
The softness ratio distribution of the major component is almost equal to 
that obtained with 32-s resolution, 
showing a strong linear correlation between the softness ratio and flare intensity (Figure 10 and 12).
The minor component also shows an overall proportionality between them;
however, it has a distinctive feature showing
a much slower increase in the softness ratio, as the flare intensity increases,
than the major component (Figure 12).
The minor component often appears as very narrow, spiky peaks in the light curves (Figure 7),
and is also responsible for the increased rapid aperiodic variability in PDSs (Figure 8).
We shall call the major component the ``broad, normal'' component;
the minor component the ``narrow, spiky'' one.
With 32-s resolution, the narrow, spiky component is dominated
by the broad, normal component, so that it is impossible to identify them.

Independent of the time resolution used,
the LMC X-4 flare spectra become softer as the flare intensities increase (Figure 10, 12, and 13).
For the case of high-luminosity sources ($>$ 10$^{38}$ ergs s$^{-1}$ for example),
the accretion column is expected to be optically thick to its own X-ray radiation 
and to re-radiate in soft X-rays (Basko \& Sunyaev 1976).
Because the number of reprocessings (i.e., scatterings/absorptions) of photons
are expected to be proportional to the flare intensities (i.e, the accretion rates),
it seems natural to attribute the origin of the broad, normal component 
to the increased optical depths of the accretion columns during the flares,
which can explain the spectral softening along with the flare intensities.
(See Basko \& Sunyaev [1976] for detailed mechanisms responsible for the spectral softening.
We assume here that the LMC X-4 flares are caused by the abruptly increased accretion rates
possibly caused by an accretion disk instability [see \S4.1].)

This mechanism can also account for the proportionality between the 
softness ratios and flare intensities of the narrow, spiky component. 
However, it is not readily capable of explaining the much slower increase 
in the softness ratios as the flare intensities increase (Figure 12),
indicating that a different mechanism is responsible for it.
We propose here that density inhomogeneity in the accretion columns onto the neutron star polar cap 
during the flares may be responsible for the observed softness ratio distributions.
If the accretion column is inhomogeneous, it is possible that some photons escape 
it without suffering as many reprocessings as expected by the flare intensities.
As indicated by the increase of the rapid aperiodic variability (Figure 9),
the inhomogeneity in the accretion columns is most likely proportional to the flare intensities,
suggesting that the probability of the leakage of photons is higher for the intenser flares.
In this case, the photons from more intense flares are expected to have
a greater chance of leaking through the inhomogeneous accretion columns.
Therefore, while the overall softness ratios are still proportional to the flare intensities, 
the intense flares are expected to show a slower increase in the softness ratio 
as the flare intensities increase
-- consistent with the observed softness ratio distributions (Figure 10 and 12).
In contrasts with the large scatter in the softness ratios of photons having
similar total photon count rates (Figure 10), 
the statistically averaged softness ratios show a well-defined systematic behavior (Figure 12).
This is also consistent with the argument given above,
because it is based on the probability of the leakage of photons,
which can best appear as a statistical behaviour.
In conclusion, the narrow, spiky peaks in the light curves (Figure 7)
seem to represent the photons leaking 
through the inhomogeneous accretion columns;
the rapid aperiodic variability proportional to the flare intensity (Figures 8 and 9) 
is most likely related to the scale of the density inhomogeneity in the accretion columns.

For bright X-ray pulsars, the previous studies have shown that 
low-density ``photon bubbles" form at the settling mound of 
the accretion column above a polar cap (Arons 1992),
because the relatively heavy accretion column is supported against gravity by radiation.
If the X-ray radiation is strong enough ($L$ $\ge$ 10$^{36}$ ergs s$^{-1}$),
a radiative shock forms at the bottom of the accretion column 
(Basko \& Sunyaev 1976; Wang \& Frank 1981; White et al. 1995).
Numerical studies (Klein et al. 1996a) have shown a variation in the number of photon bubbles
if accretion continues onto the neutron star polar cap, 
resulting in quasi-periodic variations in the luminosity
by changing the efficiency of radiation transport in the accretion column.
These are known as the ``photon-bubble oscillations" (PBOs).
Klein et al. (1996b) have found the relation,
$\nu_{\rm PBO}$ $\propto$ $L^{-0.617}$, between the frequency of the
PBOs and luminosity in the numerical simulation for the narrow polar cap model, 
assuming $\theta_{\rm c}$ = 0.1, $B$ = 3 $\times$ 10$^{12}$ G, and 
$L$ = (0.07--2) $\times$ 10$^{38}$ ergs s$^{-1}$ 
for the polar cap radius, the neutron star magnetic field,
and the luminosity, respectively.
This narrow polar cap model predicts fast ($>$ 1000 Hz) PBOs.
Due to the increased photon diffusion time, 
Klein et al. (1996b) have also found that the frequency of PBOs
decreases if the polar cap size increases.
According to their results on the larger polar cap model ($\theta_{\rm c}$ = 0.4),
$\sim$85 Hz PBOs are possible for an X-ray pulsar
with $L$ = 3 $\times$ 10$^{37}$ ergs s$^{-1}$ and $B$ = 3 $\times$ 10$^{12}$ G.
Because no previous study is directly applicable
to the extreme conditions of the LMC X-4 flares studied in this paper
($L$ $\simeq$ 10$^{39}$ erg s$^{-1}$; $B$ $\simeq$ 10$^{13}$ G),
we extended the results of Klein et al. (1996b) to the LMC X-4 flares.
If we use the $\nu_{PBO}$ $\propto$ $L^{-0.617}$ scaling law obtained from the
narrow polar cap model and $\nu_{PBO}$ = 85 Hz from the larger polar cap
model of $L$ = 3 $\times$ 10$^{37}$ ergs s$^{-1}$,
the expected PBO frequency of the LMC X-4 flares is $\sim$10 Hz for $L$ $\simeq$ 10$^{39}$ ergs s$^{-1}$,
comparable to the observed frequencies of the aperiodic variabilities
during the flares (Figure 8). We need a more through numerical simulation directly
applicable to the LMC X-4 flares to investigate the possible relations between
the rapid aperiodic variability discovered in this paper and PBOs.

\subsubsection{Origin of the Super-Eddington Radiation of the Flares}

The flare luminosities obtained in this paper are often well in excess of the Eddington luminosity of the
1.4 $M_{\odot}$ neutron star in LMC X-4, $\sim$2 $\times$ 10$^{38}$ ergs s$^{-1}$.
However, the Eddington luminosity is the upper limit that a
steady, isotropic, homogeneous, and fully ionized atmosphere can support,
which is not the case for LMC X-4 primarily because that most emission is through the neutron star polar cap.
The size of the neutron star polar cap is expected to be much smaller 
than the total neutron star surface (i.e., $\leq$ 10 \%). 
Therefore, if all the observed emission is through the polar cap, 
this indicates that sometimes LMC X-4 flare luminosities 
are above the Eddington limit by more than two orders of magnitude.
However, because a significant fraction of the emission may be transported as 
fan beams through the sides of accretion columns above the polar cap 
(due to the increased optical depths along the direction of the magnetic field),
it is not clear if the polar cap emission can really decrease the Eddington limit
by such magnitude.
On the other hand, the photon scattering cross-section in a strongly
magnetized plasma has been known to be considerably reduced 
along the magnetic field lines for the photons of which energy is much lower than the cyclotron energy 
(White, Nagese, \& Parmar 1995, and references therein).
According to the results of Paczy$\rm \acute n$ski (1992),
the strong magnetic field ($\sim$10$^{13}$ G) of LMC X-4 can render its X-ray luminosity 
up to $\sim$50 times of the Eddington limit
at the neutron star surface by reducing Thomson and Compton cross sections.
However, this effect is expected to be substantially diminished during the LMC X-4 flares. 
This is because, during the flares, radiative shocks can easily form at the bottom of the accretion columns,
supporting the accretion columns above the neutron star polar cap
(Basko \& Sunyaev 1976; Wang \& Frank 1981; White et al. 1995),
which would significantly decrease the effect of the magnetic field on the photon cross sections 
(note that the magnetic field is decaying $1 / r^3$).
Therefore, we regard the LMC X-4 flares as super-Eddington phenomena.

What then is the origin of the super-Eddington radiation of the LMC X-4 flares ?
We suggest that the inhomogeneous accretion columns during the flares, 
which were already proposed to explain the softness ratio distributions and rapid aperiodic variability (\S4.2.1), 
are also responsible for the super-Eddington radiation.
Recently, some authors have proposed that a magnetized, radiation pressure-dominated
atmosphere (and an accretion disk) is susceptible to a density inhomogeneity
(e.g., Shaviv 1998, 2000; Begelman 2001, 2002), 
most likely caused by the non-linear development 
of a ``photon-bubble instability'' (Arons 1992; Gammie 1998).
In this mechanism, once the matter is trapped in the magnetic field by radiation pressure,
the  magnetic field prevents the trapped matter from spreading sideways, maintaining the density inhomogeneity.
The net flux through this inhomogeneous atmosphere can easily exceed the
Eddington limit by a large factor primarily through the low-density (= low-opacity) regions
(Shavie 1998, 2000; Begelman 2001).
We note that the LMC X-4 flares satisfy the conditions for the development 
of the density inhomogeneity in the accretion columns easily
owing to their high luminosities and the strong magnetic field of the source. 
In addition, because a radiative shock can readily develop 
above the polar cap due to the high luminosities of the LMC X-4 flares,
the ``photon bubbles'' are expected to form at the bottom of the accretion columns during the flares,
which can be a seed for the proposed density inhomogeneity.

\subsection{Pulse Profiles}

During the LMC X-4 flares, the accretion columns are expected to be highly opaque.
In this case, one might expect that a significant fraction of the flare emission 
is transported via the fan beams in the direction perpendicular to the magnetic field
due to the relatively smaller optical depths in this direction than the direction
parallel to the magnetic field.
The increased fan beams then can result in substantial
gravitational bendings of the X-ray photons from the neutron star polar cap 
(e.g., Brainerd \& M$\acute e$sz$\acute a$ros 1991), 
which may be responsible for the broad pulse profiles of the LMC X-4 flares.
On the other hand, the simple sinusoidal, in-phase pulse profiles 
may indicate that pulse profiles are simply dominated by one mechanism,
i.e., the flares, probably at one magnetic pole.

\section{Summary and Conclusion}

In this paper, we have analyzed two extensive observations of the
LMC X-4 flares made with RXTE. We summarize our main results as follows.
(1) The flare light curves reveal the existence of the flat precursors preceding the flares,
with the complex activity of the LMC X-4 flares.
(2) The FWHM and mean fluence of the flares are
68 $\pm$ 31 s and (3.1 $\pm$ 2.9) $\times$ 10$^{40}$ ergs, respectively.
(3) The pulse profiles during the flares are broad, sinusoidal,
and the pulsed fractions increase with the flare intensities.
(4) With 0.0625-s resolution, narrow, spiky peaks appear in 
the light curves, so does the rapid aperiodic variability in PDSs.
Both intensify along with the flare intensities.
(6) The softness ratios and spectral fits show that the flare spectra
soften as the intensities increase.
(7) With 0.0625-s resolution, while the photons having count rates
less than 300 counts/sec/PCU show the softness ratio distributions 
almost equal to those obtained with 32-s resolution, 
the photons having larger photon count rates 
show a much slower increase in the softness ratios as the flare intensities increase. 
(8) The peak luminosity of the flares reaches up to 
$\sim$2.1 $\times$ 10$^{39}$ ergs s$^{-1}$, well above the Eddington limit
for an 1.4 $M_{\odot}$ neutron star.

We find it difficult to explain the observed properties of
the LMC X-4 flares in terms of the thermonuclear instability on the neutron star surface 
or the viscous instability at the inner accretion disk.
We consider the possibility that the plasma Rayleigh-Taylor instability 
in the accretion disk may be responsible for the flares.
We propose that increased optical depths and density inhomogeneity
in the accretion columns onto the neutron star polar cap 
during the LMC X-4 flares are responsible for the observed properties, 
including the super-Eddington luminosities of the flares. 
If confirmed, this is the first observational evidence for the inhomogeneous 
atmosphere responsible for super-Eddington radiation. 
The broad pulse profiles of the LMC X-4 flares
may be caused by substantial gravitational bending of the flare emission
transported via fan beams.

\acknowledgments
We would like to thank the anonymous referee for his (or her) comments
which substantially improved this paper.
We also would like to thank Ron Remillard for his help on the RXTE data analyses.
D.-S. M. acknowledges Akiko Shirakawa for her comments.
This research has made use of data obtained from the {\it
High Energy Astrophysics Science Archive Research Center}
provided by NASA's Goddard Space Flight Center.
D.-S. M. is supported by NSF grant AST-9986898,
and S. S. E. is supported in part by an NSF Faculty Early Careeer Development award (NSF-9983830).
I. M. W. is supported by the NASA grant NAG-5-8356.

\clearpage
\begin{deluxetable}{lrrrrrr}
\tablecolumns{7}
\tablewidth{0pt}
\tablecaption{Parameters of the LMC X-4 Flare Peaks \label{tbl-1}}
\tablehead{
\colhead{Flare} & \colhead{position} & \colhead{peak\tablenotemark{a}} & \colhead{FWHM}   & \colhead{dof} & \colhead{$\chi^2_{\nu}$} & fluence\tablenotemark{b}                  \\ 
\colhead{     } & \colhead{(sec)}    & \colhead{               }       & \colhead{(sec)}  & \colhead{   } & \colhead{              } & {(10$^{39}$ ergs)} \\  }
\startdata
FL1 & 2435 & 121 & 123 & 16 & 3.5 &    60 \\
FL1 & 3053 & 134 & 48 & 6 & 4.0 &    23 \\
FL1 & 3181 & 147 & 78 & 10 & 3.5 &    42 \\
FL1 & 3762 & 142 & 38 & 5 & 3.2 &    16 \\
FL1 & 3974 & 154 & 66 & 8 & 3.5 &    34 \\
FL1 & 4118 & 97 & 74 & 9 & 0.9 &    30 \\
FL2 & 1184 & 294 & 84 & 11 & 12.9 &    80 \\
FL2 & 1354 & 351 & 50 & 6 & 6.3 &    46 \\
FL2 & 1533 & 218 & 152 & 19 & 6.9 &   112 \\
FL2 & 1723 & 126 & 76 & 10 & 1.3 &    40 \\
FL2 & 1843 & 152 & 101 & 12 & 1.6 &    59 \\
FL2 & 1998 & 106 & 29 & 4 & 0.1 &    10 \\
FL2 & 2673 & 190 & 89 & 11 & 2.6 &    58 \\
FL2 & 2868 & 200 & 94 & 12 & 4.9 &    68 \\
FL2 & 3218 & 101 & 30 & 4 & 1.5 &    11 \\
FL2 & 4212 & 142 & 44 & 6 & 4.2 &    21 \\
FL2 & 4379 & 172 & 42 & 6 & 2.1 &    21 \\
FL2 & 4508 & 146 & 57 & 8 & 1.8 &    31 \\
FL3 & 3095 & 232 & 38 & 5 & 1.5 &    27 \\
FL3 & 3239 & 355 & 37 & 5 & 18.8 &    37 \\
FL3 & 3449 & 265 & 65 & 8 & 5.1 &    58 \\
FL3 & 3882 & 162 & 35 & 4 & 0.7 &    20 \\
FL3 & 4055 & 187 & 70 & 9 & 1.6 &    47 \\
FL3 & 4172 & 137 & 69 & 8 & 1.2 &    36 \\
FL3 & 4301 & 97 & 96 & 12 & 0.6 &    42 \\
FL3 & 4510 & 100 & 86 & 11 & 1.6 &    36 \\
FL4 & 1772 & 606 & 111 & 14 & 26.0 &   187 \\
FL4 & 3548 & 293 & 113 & 14 & 7.6 &   113 \\
FL5 & 1340 & 282 & 36 & 4 & 16.9 &    14 \\
FL5 & 1438 & 424 & 29 & 4 & 30.7 &    11 \\
FL5 & 1592 & 464 & 15 & 2 & 0.1 &     5 \\
FL5 & 1787 & 275 & 24 & 3 & 5.8 &    11 \\
FL5 & 2010 & 395 & 87 & 11 & 11.7 &    28 \\
FL5 & 2283 & 254 & 120 & 16 & 4.4 &    29 \\
FL5 & 2471 & 321 & 13 & 1 & 0.2 &     2 \\
FL6 &  539 & 193 & 52 & 6 & 9.2 &     7 \\
FL6 & 1555 & 145 & 45 & 6 & 0.2 &     6 \\
FL6 & 2173 & 95 & 79 & 10 & 0.9 &     9 \\
FL7 & 2889 & 92 & 84 & 11 & 1.4 &    14 \\
FL7 & 3098 & 109 & 68 & 9 & 1.2 &     9 \\
FL7 & 3414 & 98 & 65 & 8 & 0.5 &    10 \\
FL7 & 3490 & 95 & 35 & 4 & 0.4 &     6 \\
FL7 & 3572 & 105 & 52 & 6 & 0.9 &     5 \\
FL7 & 3778 & 118 & 85 & 11 & 1.4 &    16 \\
FL7 & 4024 & 143 & 40 & 5 & 1.4 &     5 \\
FL7 & 4222 & 116 & 34 & 4 & 0.3 &     5 \\
FL7 & 4406 & 101 & 32 & 4 & 2.6 &     3 \\
FL8 &  598 & 90 & 163 & 21 & 1.0 &    19 \\
FL8 & 1261 & 125 & 53 & 6 & 3.5 &     7 \\
FL8 & 1772 & 107 & 88 & 11 & 1.6 &     9 \\
FL8 & 2317 & 108 & 58 & 7 & 1.4 &     7 \\
FL8 & 2880 & 92 & 126 & 15 & 0.6 &    15 \\
FL8 & 4124 & 90 & 52 & 6 & 1.0 &    16 \\
FL9 &  493 & 105 & 82 & 10 & 1.5 &    27 \\
FL9 & 1123 & 98 & 93 & 12 & 1.2 &    23 \\
FL9 & 1549 & 111 & 94 & 12 & 2.2 &    33 \\
FL9 & 1769 & 93 & 72 & 9 & 0.9 &    25 \\
FL9 & 2230 & 119 & 55 & 7 & 2.8 &    12 \\
FL9 & 2673 & 117 & 89 & 11 & 2.9 &    32 \\
FL9 & 3556 & 102 & 105 & 14 & 1.1 &    28 \\
FL9 & 4196 & 121 & 60 & 8 & 0.8 &    14 \\
FL9 & 4503 & 128 & 55 & 7 & 2.4 &    15 \\
FL10 & 2808 & 103 & 51 & 7 & 1.5 &    29 \\
FL10 & 3261 & 136 & 58 & 7 & 1.9 &    37 \\
FL10 & 3600 & 135 & 67 & 9 & 1.7 &    39 \\
FL10 & 4009 & 155 & 54 & 7 & 1.1 &    23 \\
FL10 & 4461 & 147 & 51 & 6 & 4.6 &    24 \\
FL10 & 4565 & 95 & 78 & 10 & 1.2 &    23 \\
\enddata 
\tablenotetext{a}{In unit of counts/sec/PCU.}
\tablenotetext{b}{Caculated at 2.5--25 keV.}
\end{deluxetable} 

\clearpage
\begin{deluxetable}{rlll}
\tablecolumns{4}
\tablewidth{0pt}
\tablecaption{Correlation Coefficients and Slopes of the Lines in Figures 10 and 11\label{tbl-2}}
\tablehead{
\colhead{Resolution}  & \colhead{All data}                     & \colhead{Faint group}                 & \colhead{Bright group}   }
\startdata
32 s\tablenotemark{a}      & 0.95, 1.4  $\times$ 10$^{-2}$ (240)    & \nodata                               &  \nodata \\ 
0.0625 s\tablenotemark{a}  & 0.30, 7.8  $\times$ 10$^{-3}$ (122880) & 0.19, 1.1 $\times$ 10$^{-2}$ (110592) &  --0.01,  --2.4  $\times$ 10$^{-4}$ (1228)  \\ 
0.0625 s\tablenotemark{b}  & 0.20(13), 1.0(0) $\times$ 10$^{-2}$    & 0.09(6), 9.8(0) $\times$ 10$^{-3}$    & 0.25(4), 8.2(6) $\times$ 10$^{-3}$  \\
\enddata
\tablenotetext{a}{From the real data. The numbers in parentheses are the number of data points.}
\tablenotetext{b}{From the Monte Carlo simulation. The numbers in parentheses correspond to 1$\sigma$ deviations at the last quoted digit.}
\end{deluxetable}

\clearpage
\begin{deluxetable}{lrrr}
\tablecolumns{4}
\tablewidth{0pt}
\tablecaption{Luminosity and Fluence of FL1--10 \label{tbl-3}}
\tablehead{
\colhead{Name}            & \colhead{precursor}                   & \colhead{peak}                      & \colhead{fluence\tablenotemark{a}} \\ 
\colhead{ }               & \colhead{(10$^{38}$ ergs s$^{-1}$)}   & \colhead{(10$^{38}$ ergs s$^{-1}$)} & \colhead{(10$^{42}$ ergs)} }
\startdata
FL1                       & \nodata              & 6.1 $\pm$ 1.0        &   0.7 $\pm$ 0.2   \\
FL2                       & 1.8 $\pm$ 0.5        & 13.0 $\pm$ 1.6       &   1.4 $\pm$ 0.3   \\
FL3                       & 1.9 $\pm$ 0.4        & 12.9 $\pm$ 1.7       &   1.1 $\pm$ 0.3   \\
FL4                       & 2.0 $\pm$ 0.3        & 21.4 $\pm$ 2.6       &   1.9 $\pm$ 0.4   \\
FL5                       & \nodata              & 4.8 $\pm$ 0.2        &   0.8 $\pm$ 0.3 \\
FL6                       & \nodata              & 2.3 $\pm$ 0.3        &   0.5 $\pm$ 0.2  \\
FL7                       & \nodata              & 1.9 $\pm$ 0.3        &   0.5 $\pm$ 0.2 \\
FL8                       & \nodata              & 1.7 $\pm$ 0.4        &   0.5 $\pm$ 0.3   \\
FL9                       & \nodata              & 3.9 $\pm$ 0.8        &   1.0  $\pm$ 0.3   \\
FL10                      & \nodata              & 6.1 $\pm$ 0.9        &   1.1  $\pm$ 0.3  \\
\enddata
\tablenotetext{a}{Including the normal state.}
\tablecomments{The errors correspond to 1 $\sigma$ uncertanties, and the energy range is 2.5--25 keV}
\end{deluxetable}

\clearpage
\begin{deluxetable}{lrrrcr}
\tablecolumns{6}
\tablewidth{0pt}
\tablecaption{Best-fit Parameters of the Brightest Peaks in FL1--10 \label{tbl-4}}
\tablehead{
\colhead{Name}            & \colhead{$N_{\rm H}$}                 & \colhead{$\alpha$}    &  \colhead{$kT$}   & \colhead{flux\tablenotemark{a}}                      & \colhead{$\chi^2_{\nu}$}  \\
\colhead{ }               & \colhead{(10$^{22}$ cm$^{-2}$)}   & \colhead{        }    &  \colhead{(keV)}  & \colhead{(10$^{-9}$ ergs s$^{-1}$)}   & \colhead{ }               }
\startdata
FL1                       & 2.4$^{-1.3}_{+1.2}$  & 1.8$^{-0.2}_{+0.2}$   & 2.3$^{-0.3}_{+0.5}$   & 2.0 $\pm$ 0.3   &  0.9     \\
FL2                       & 3.6$^{-1.0}_{+1.0}$  & 2.5$^{-0.2}_{+0.1}$   & 1.7$^{-0.1}_{+0.1}$   & 4.3 $\pm$ 0.5   &  1.2     \\
FL3                       & 2.5$^{-0.9}_{+1.1}$  & 2.4$^{-0.2}_{+0.1}$   & 1.6$^{-0.1}_{+0.2}$   & 4.2 $\pm$ 0.6   &  1.4      \\
FL4                       & 4.2$^{-0.8}_{+1.0}$  & 3.0$^{-0.2}_{+0.1}$   & 1.4$^{-0.1}_{+0.1}$   & 7.2 $\pm$ 0.9   &  1.1      \\
FL5                       & 1.8\tablenotemark{b} & 2.5$^{-0.1}_{+0.1}$   & 1.5$^{-0.1}_{+0.1}$   & 1.6 $\pm$ 0.1   &  1.6      \\
FL6                       & 1.8\tablenotemark{b} & 1.9$^{-0.1}_{+0.4}$   & 2.3$^{-0.2}_{+1.3}$   & 0.8 $\pm$ 0.1   &  1.4      \\
FL7                       & 1.8\tablenotemark{b} & 2.9$^{-0.5}_{+0.4}$   & 4.2$^{-0.4}_{+0.3}$   & 0.6 $\pm$ 0.1   &  1.2      \\
FL8                       & 1.0$^{-0.9}_{+1.1}$  & 2.5$^{-0.5}_{+0.5}$   & 4.7$^{-0.4}_{+0.5}$   & 0.6 $\pm$ 0.1   &  0.8     \\
FL9                       & 1.8\tablenotemark{b} & 2.5$^{-0.6}_{+0.4}$   & 5.3$^{-0.5}_{+0.5}$   & 1.6 $\pm$ 0.3   &  1.3      \\
FL10                      & 1.1$^{-1.0}_{+1.1}$  & 1.9$^{-0.3}_{+0.4}$   & 3.9$^{-0.6}_{+0.5}$   & 2.0 $\pm$ 0.3   &  1.4      \\
\enddata
\tablenotetext{a}{In the 2.5--25 keV energy range.}
\tablenotetext{b}{The values are fixed in the fits.}
\tablecomments{The errors correspond to 1 $\sigma$ uncertanties.}
\end{deluxetable}

\clearpage
\begin{deluxetable}{ccccccccc}
\tablecolumns{9}
\tablewidth{0pt}
\tablecaption{Best-fit Parameters of the Spectral Fits \label{tbl-5}}
\tablehead{
\colhead{Parameters} & \colhead{$N_{\rm H}$}            & \colhead{$\alpha$}   & \colhead{$kT$}       & \colhead{$E_{\rm c}$}   & \colhead{$E_{\rm f}$} & \colhead{$E_{\rm G}$}   & \colhead{flux\tablenotemark{a}}               & \colhead{$\chi^2_{\nu}$}     \\
\colhead{        }   & \colhead{(10$^{22}$ cm$^{-2}$)}  & \colhead{}           & \colhead{(keV)}      & \colhead{(keV)}         & \colhead{(keV)}       & \colhead{(keV)}         & \colhead{}   & \colhead{}                  \\}
\startdata
\cutinhead{Model 1\tablenotemark{b}}
Flare Peak           &     4.8$^{-0.8}_{+1.0}$          & 3.1$^{-0.1}_{+0.2}$    & 1.4$^{-0.1}_{+0.1}$  & \nodata                 & \nodata               & 6.3$^{-0.2}_{+0.2}$     & 7.1   & 1.0 \\
Normal State         &     4.5$^{-1.5}_{+2.8}$          & 2.1$^{-0.4}_{+0.7}$    & 6.0$^{-0.5}_{+0.4}$  & \nodata                 & \nodata               & 7.2$^{-0.4}_{+0.3}$     & 7.1   & 1.4 \\
\cutinhead{Model 2\tablenotemark{c}}
Flare Peak           &     5.7$^{-0.5}_{+0.5}$          & 3.3$^{-0.2}_{+0.1}$     & \nodata              & \nodata                 & \nodata               & 5.7$^{-0.6}_{+0.3}$     & 7.2   & 1.4 \\
Normal State         &     1.9$^{-0.5}_{+0.5}$          & 0.98$^{-0.03}_{+0.03}$  & \nodata              & \nodata                 & \nodata               & 7.3$^{-0.5}_{+0.5}$     & 0.69  & 1.7  \\
\cutinhead{Model 3\tablenotemark{d}}
Flare Peak           &     2.9$^{-0.8}_{+2.0}$          & 2.0$^{-0.3}_{+0.4}$    & \nodata              & 3.6$^{-3.5}_{+2.5}$     & 8.3$^{-1.2}_{+3.6}$   & 5.9$^{-0.4}_{+0.4}$     & 7.1   & 1.0 \\
Normal State         &     0.60$^{-0.5}_{+1.2}$          & 0.81$^{-0.30}_{+0.31}$  & \nodata            & 16.1$^{-2.4}_{+2.5}$     & 27.4$^{-13.5}_{+14.6}$   & 6.3$^{-1.0}_{+0.9}$     & 0.54  & 1.5 \\
\label{table3}
\enddata
\tablenotetext{a}{In unit of of 10$^{-9}$ ergs cm$^{-2}$ s$^{-1}$.}
\tablenotetext{b}{See the Eqn. (1).}
\tablenotetext{c}{Same as the Eqn. (1), but without the black body component.}
\tablenotetext{d}{See the Eqn. (2).}
\end{deluxetable}

\clearpage
\begin{figure}
\plotone{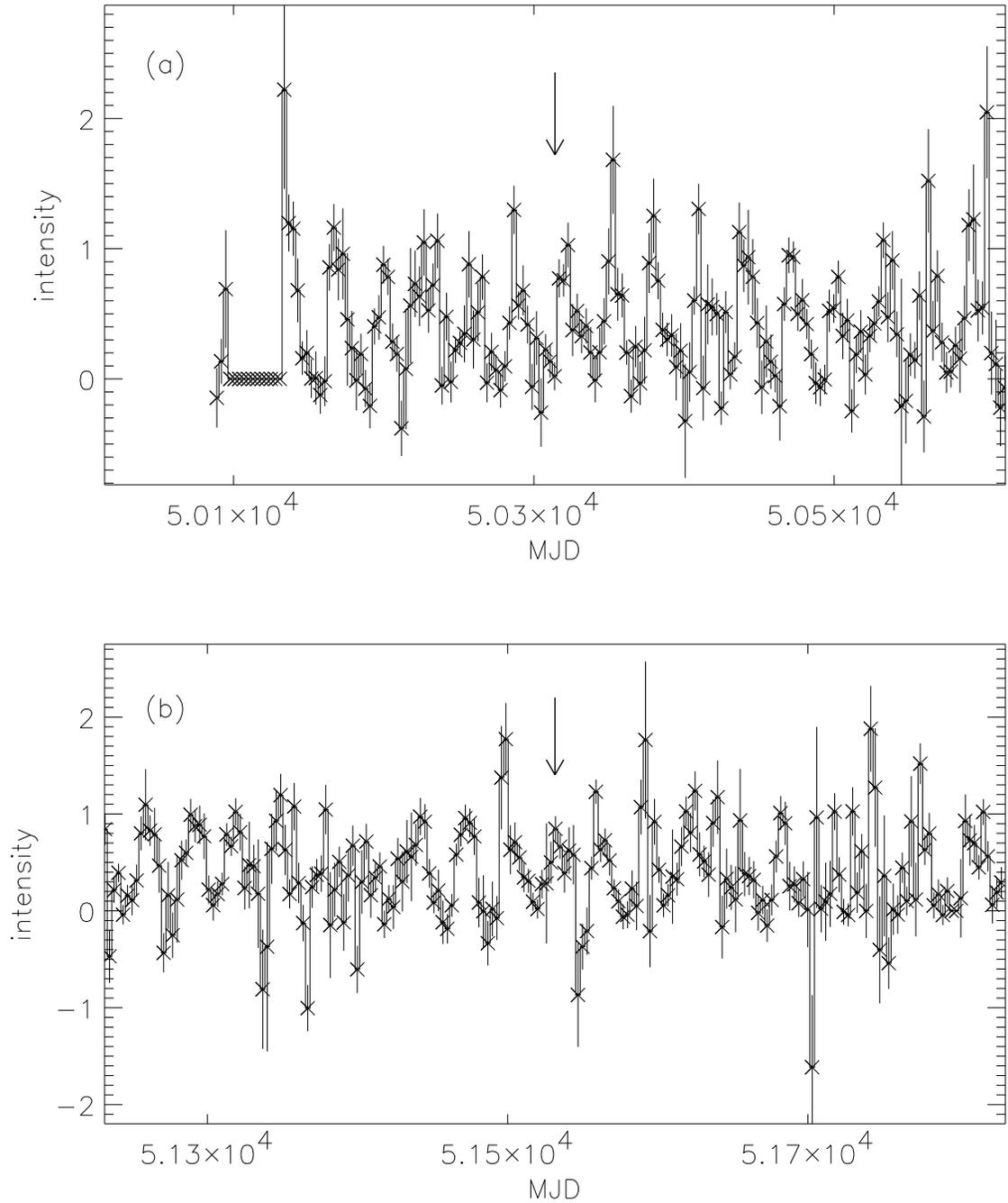}
\caption{The RXTE All Sky Monitor data of LMC X-4 around the 
two observations reported in this paper -- (a) for the 1996 August
observations and (b) for the 1999 December ones.
The arrows show the start times of the two observations, and 
the error bars correspond to 1$\sigma$ uncertainties.
\label{fig1}}
\end{figure}

\clearpage
\begin{figure}
\plotone{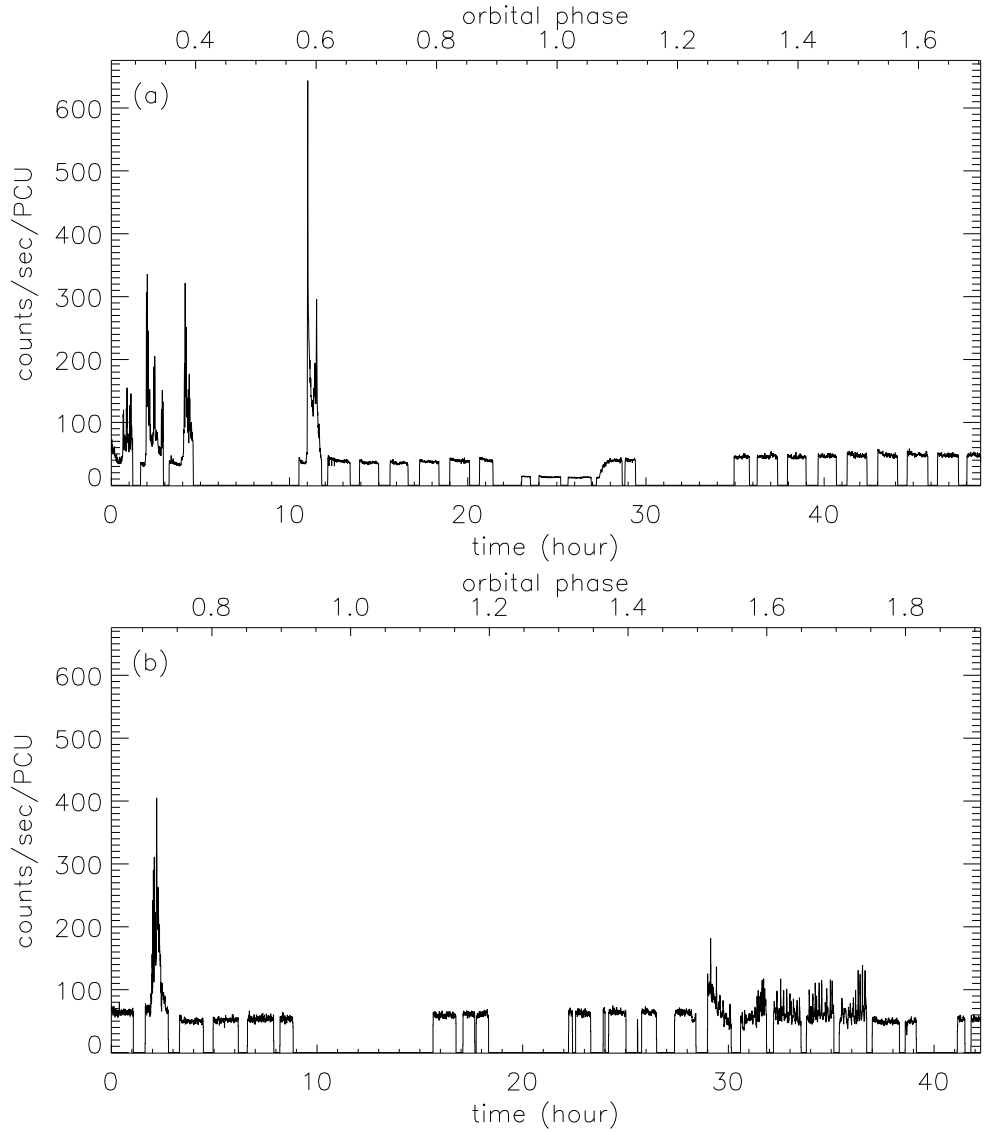}
\caption{Background-subtracted light curves of LMC X-4
in the 2--25 keV range with 32-s resolution: 
(a) for the 1996 observations and (b) for the 1999 ones.
Upper x-axes represent orbital phases of LMC X-4.
\label{fig2}}
\end{figure}

\clearpage
\begin{figure}
\plotone{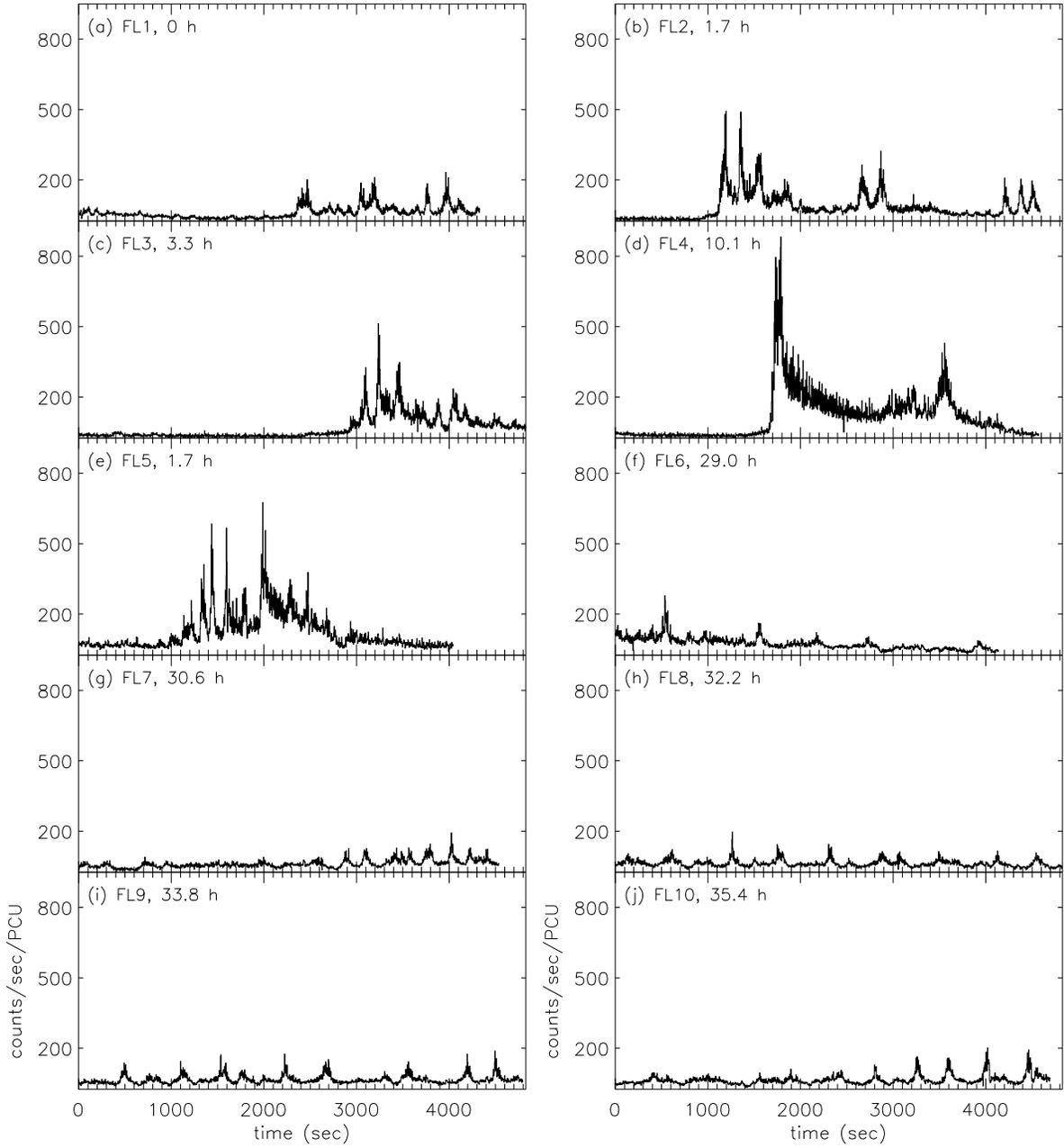}
\caption{Same as Figure 2, but only for the data segments containing the flares
with 4-s resolution. (a)--(j) are for FL1--FL10.
The inserted numbers represent the time elapsed
from the start of the two observations, 
i.e., the 1996 ones for FL1--4 and the 1999 ones for FL5--10.
\label{fig3}}
\end{figure}

\clearpage
\begin{figure}
\plotone{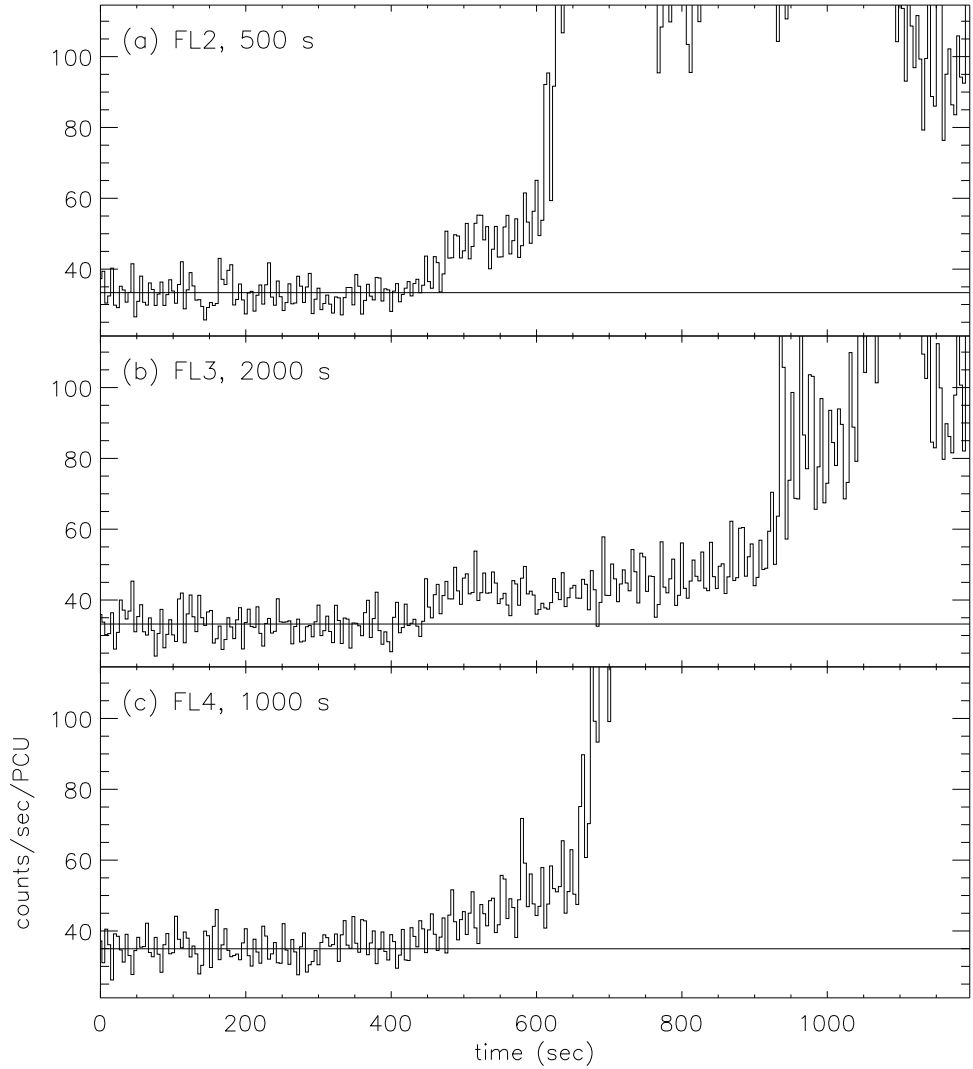}
\caption{Magnified light curves of FL2, FL3, and FL4 around the beginnings of the flares.
The inserted numbers represent the time offset from the start of each light curve in Figure 3.
\label{fig4}}
\end{figure}

\clearpage
\begin{figure}
\plotone{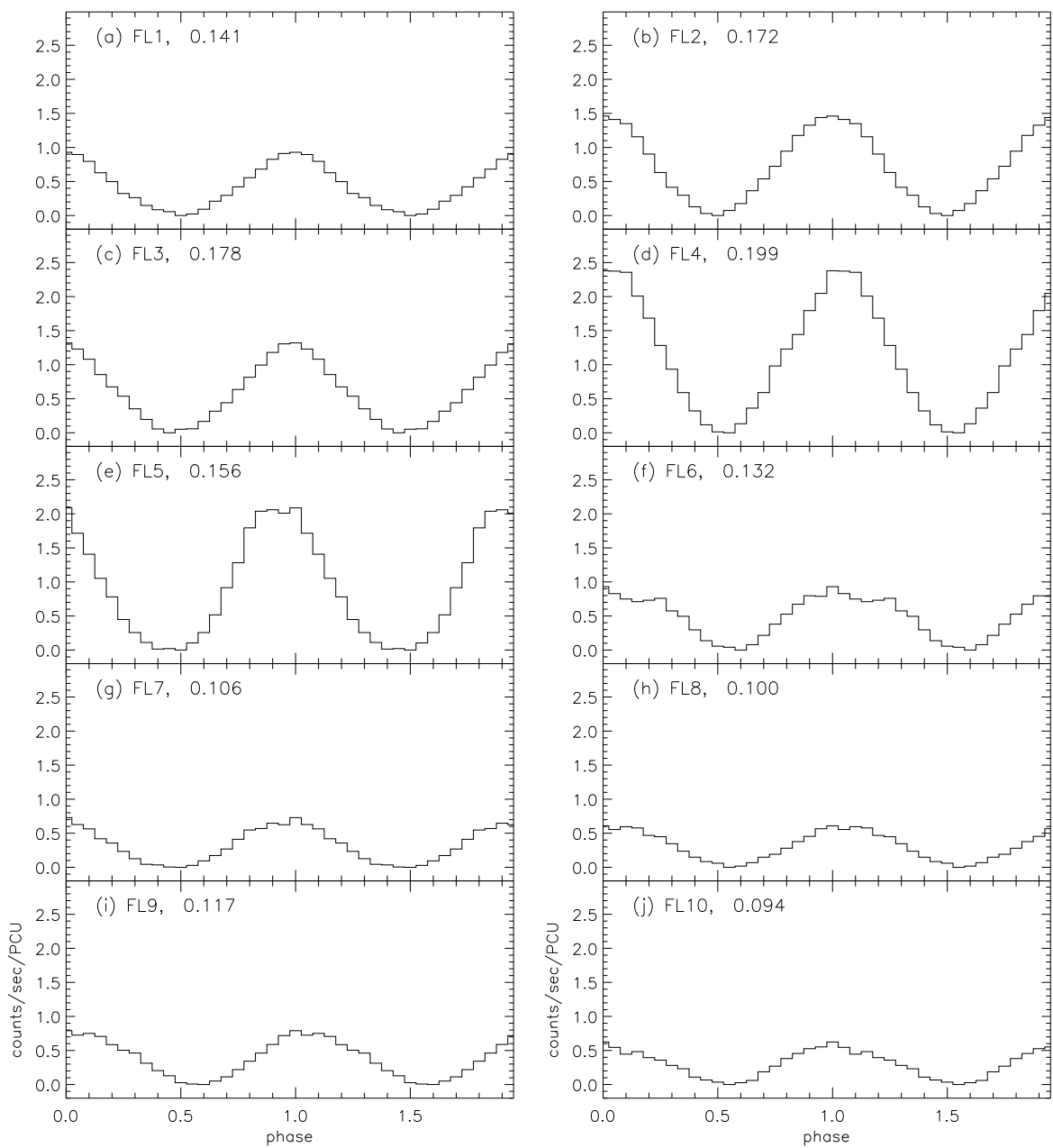}
\caption{Pulse profiles of FL1--10 in Figure 3.
The phase 0 is adjusted to have the maximum photon count rates,
and the number of bins is 20. The inserted numbers represent pulsed fractions.
\label{fig5}}
\end{figure}

\clearpage
\begin{figure}
\plotone{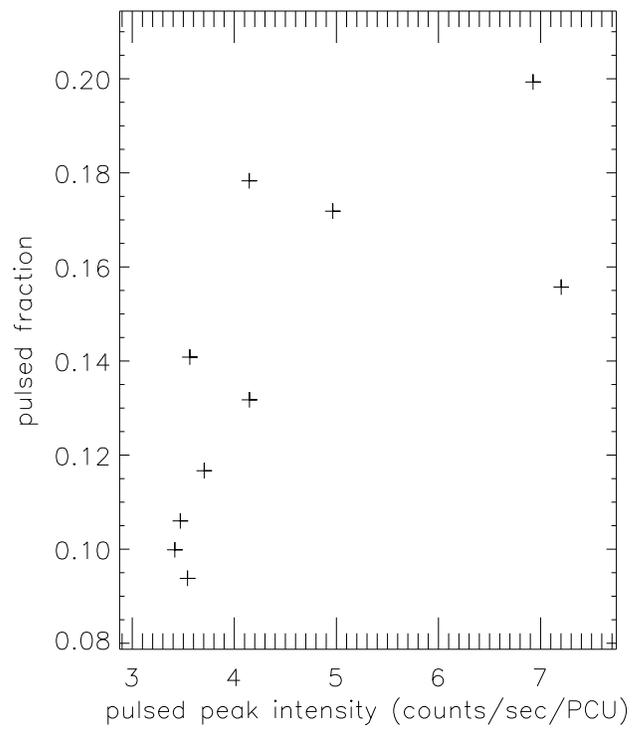}
\caption{Comparison between the pulsed fractions and the peak intensities of the pulse
profiles in Figure 5.
\label{fig6}}
\end{figure}

\clearpage
\begin{figure}
\plotone{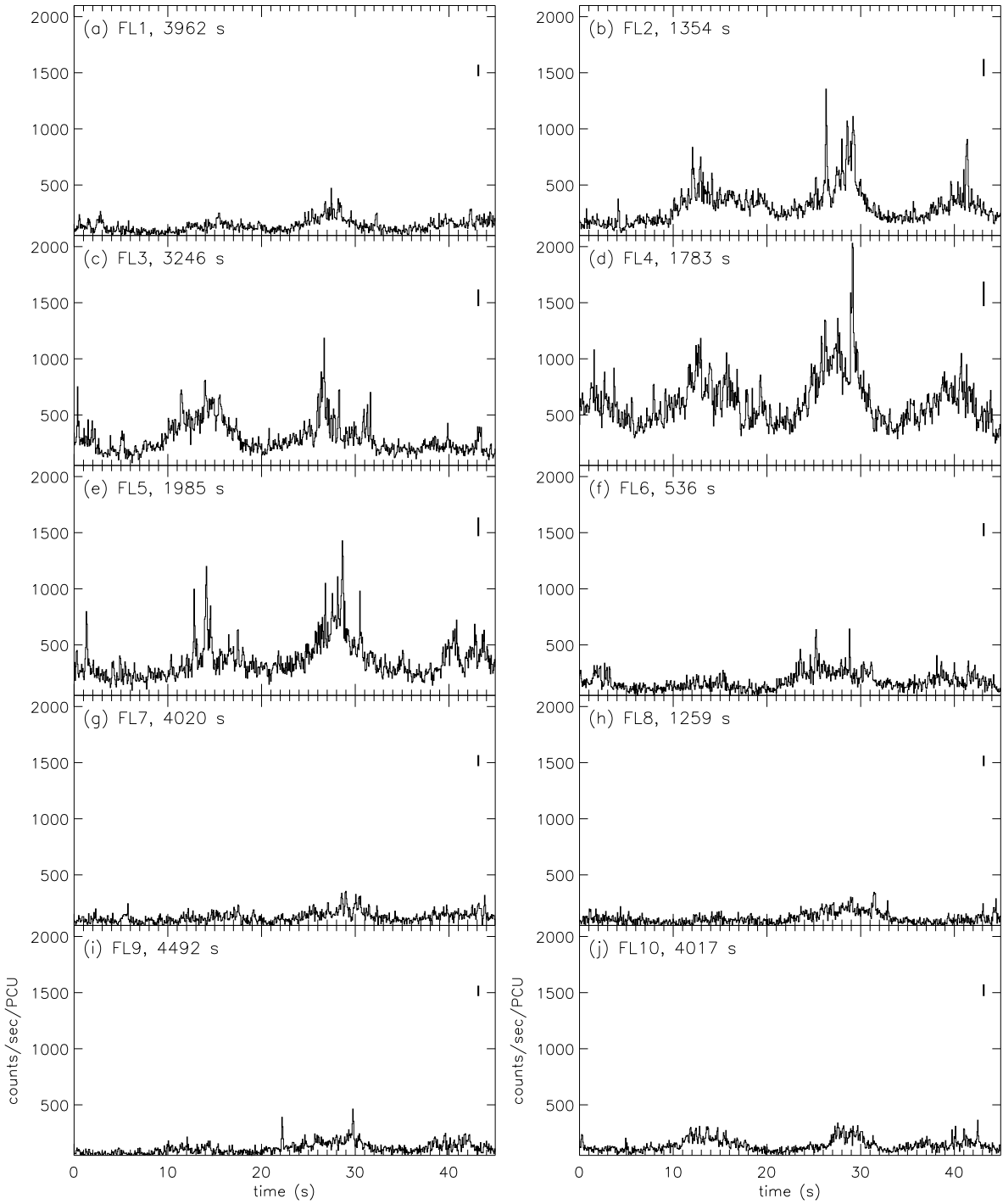}
\caption{Same as Figure 3, but only for the 45-s data segments
around the brightest peak in each light curve with 0.0625-s resolution.
The inserted number represents the time elapsed from the start of each light curve;
the length of the vertical bars represents the magnitude of the mean Poisson noise.
\label{fig7}}
\end{figure}

\clearpage
\begin{figure}
\plotone{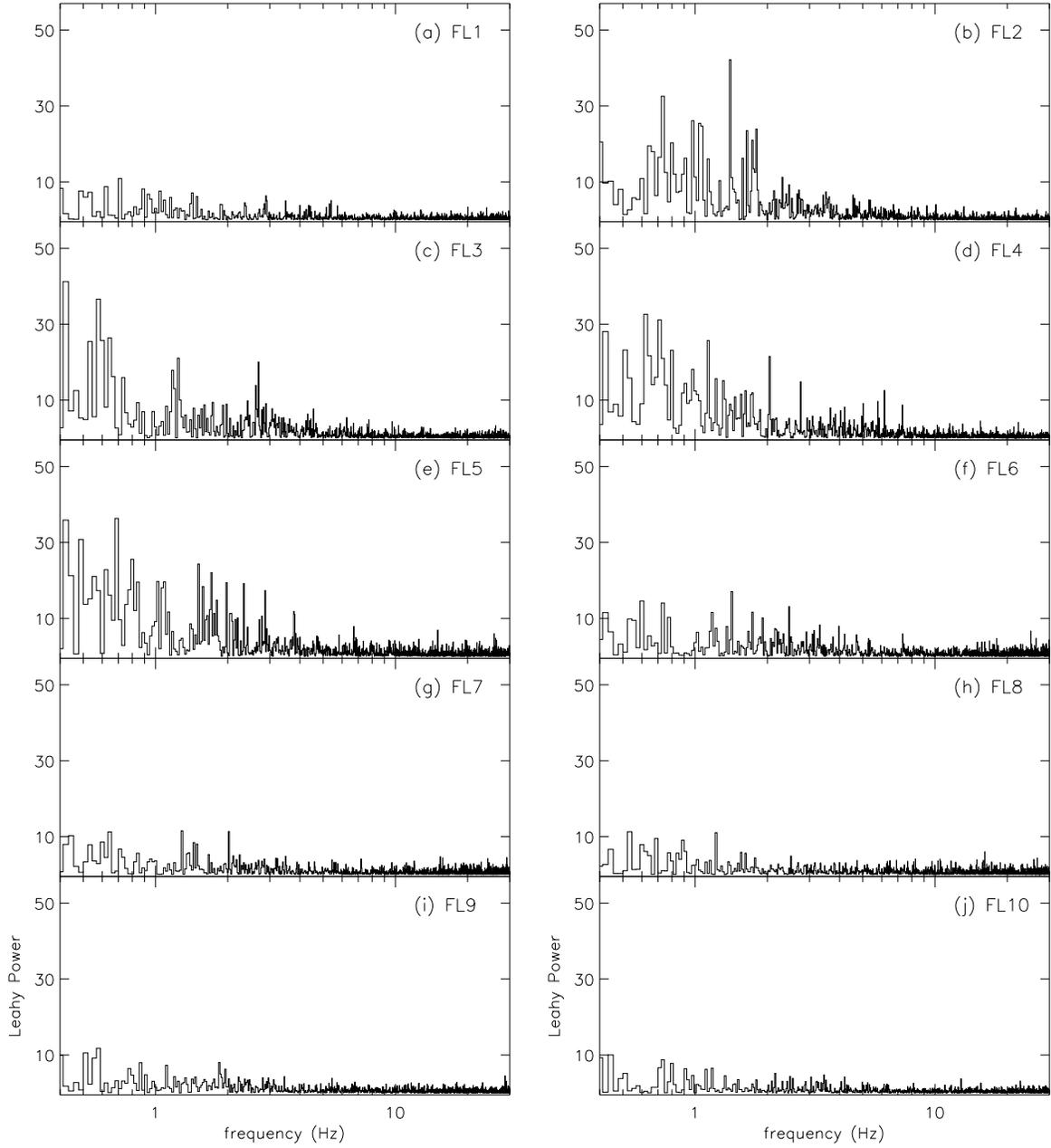}
\caption{Leahy-normalized PDSs of the 10-s segments around the brightest peak in FL1--10.
\label{fig8}}
\end{figure}

\clearpage
\begin{figure}
\plotone{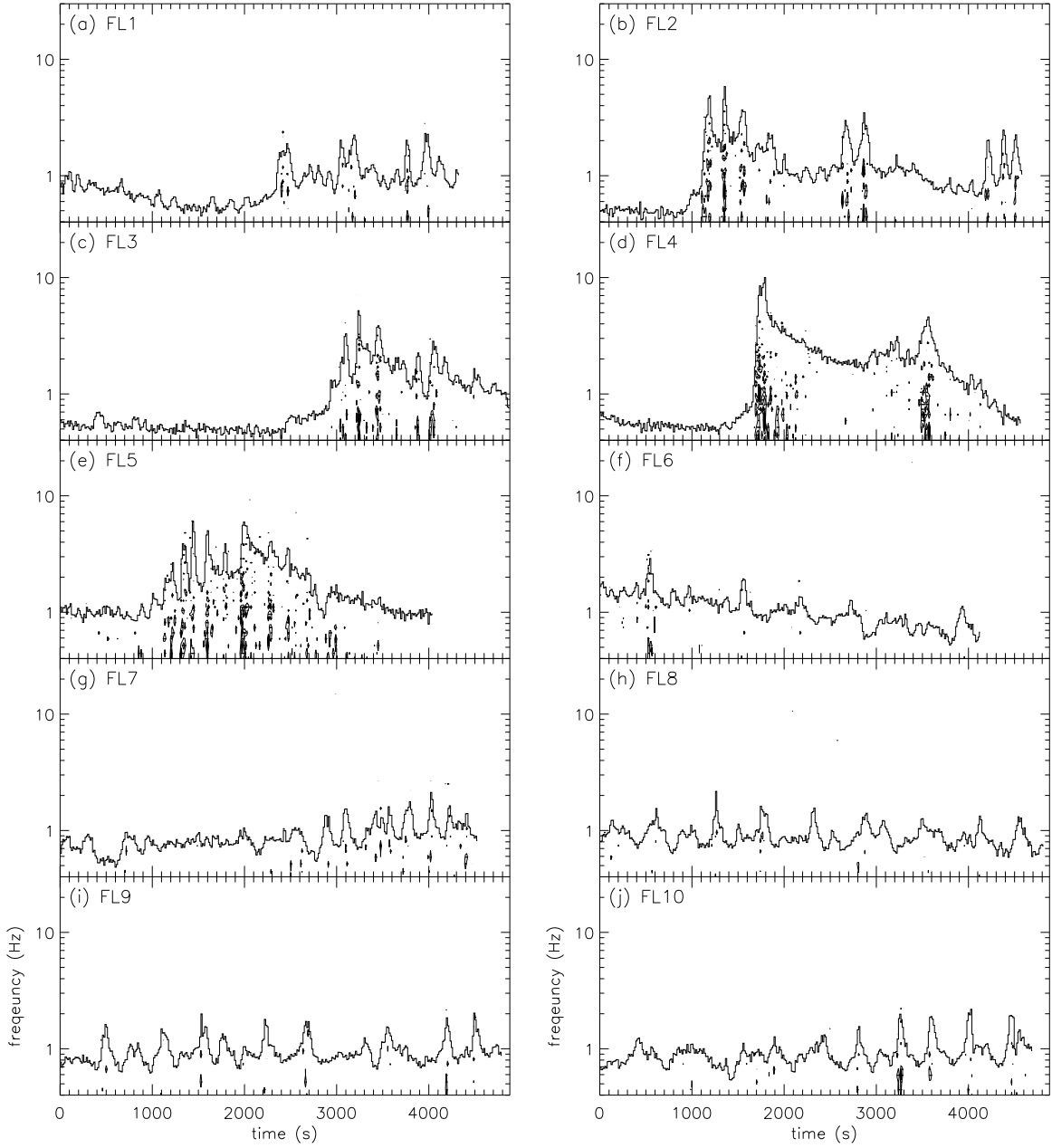}
\caption{Distributions of the integrated (over $\sim$13.5-s interval)
flare intensities (solid lines; in arbitrary logarithmic intensity scale)
and significant Leahy powers ($>$ 10; dots).
\label{fig9}}
\end{figure}

\clearpage
\begin{figure}
\plotone{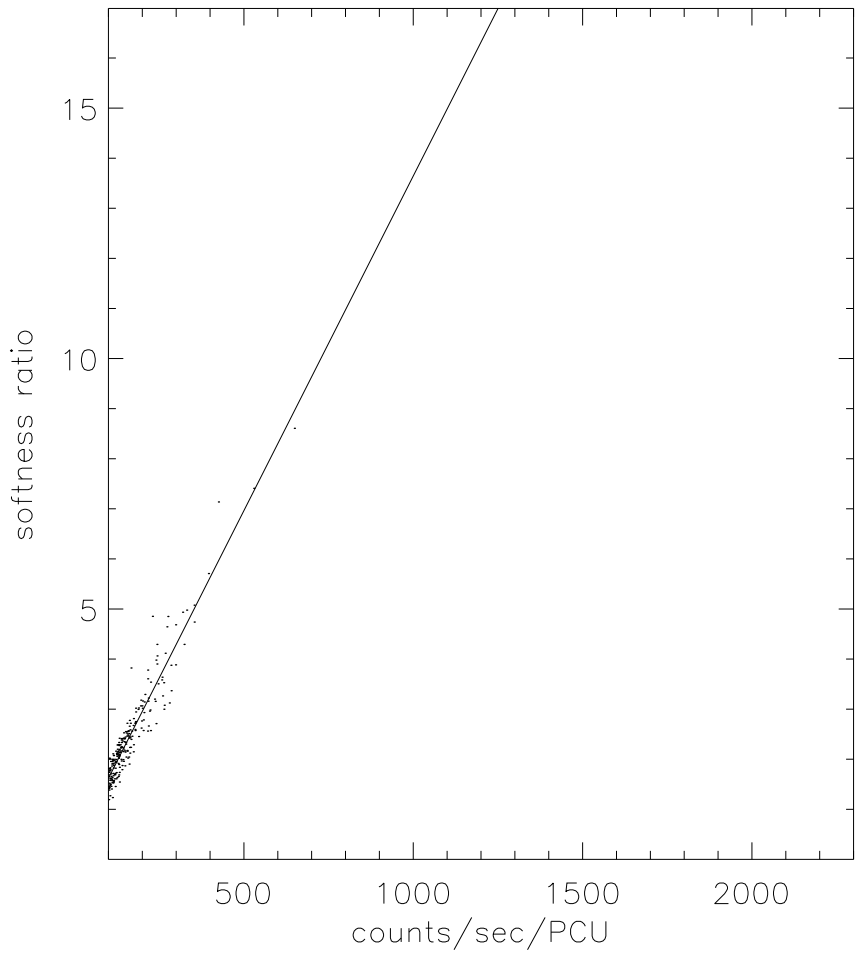}
\caption{Distribution of the softness ratios computed with 32-s resolution.
The solid line represents the linear correlation between the softness ratios
and the total photon count rates.
The line is extended to be compared with Figure 11.
\label{fig10}}
\end{figure}

\clearpage
\begin{figure}
\caption{(a) Same as Figure 10, but with 0.0625-s resoluton.
The dotted and dashed lines represent the correlations between the
softness ratios and the total photon count rates found in the lower 90 \% and
the upper 1 \% data points, respectively.
The lines are extended to the full extents of the figures 
to be seen more clearly.  
(b) Same as Figure (a), but obtained from the Monte Carlo simulation.
\label{fig11}}
\end{figure}

\clearpage
\begin{figure}
\plotone{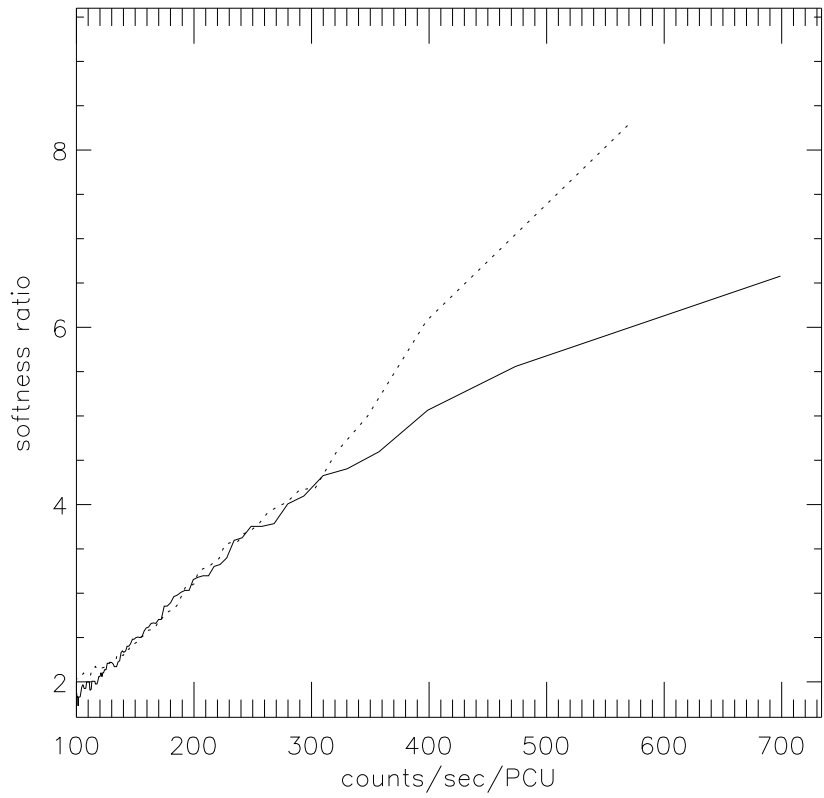}
\caption{Distributions of the softness ratios integrated in 100 total intensity bins.
The solid line is for the real data (Figure 11a); 
the dotted line is for the simulated data (Figure 11b).
\label{fig12}}
\end{figure}

\clearpage
\begin{figure}
\plotone{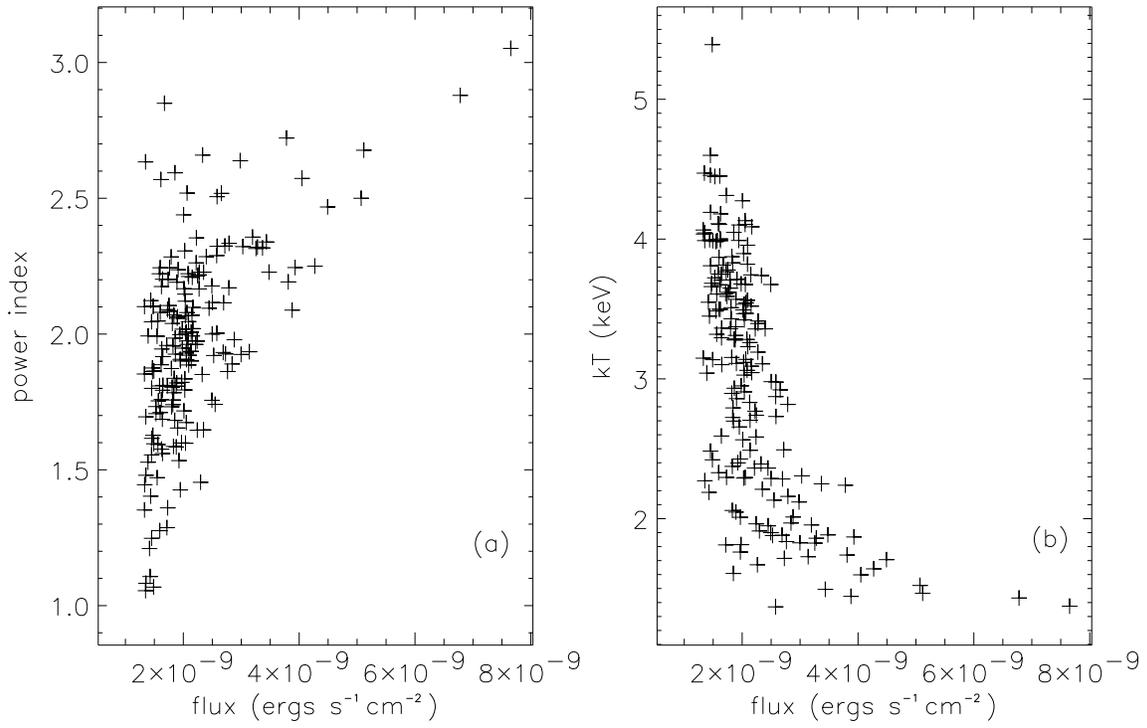}
\caption{Comparisons of the fitted power index (a) and the black body temperature (b)
of the flares with the flux at 2.5--25 keV.
\label{fig13}}
\end{figure}

\clearpage
\begin{figure}
\plotone{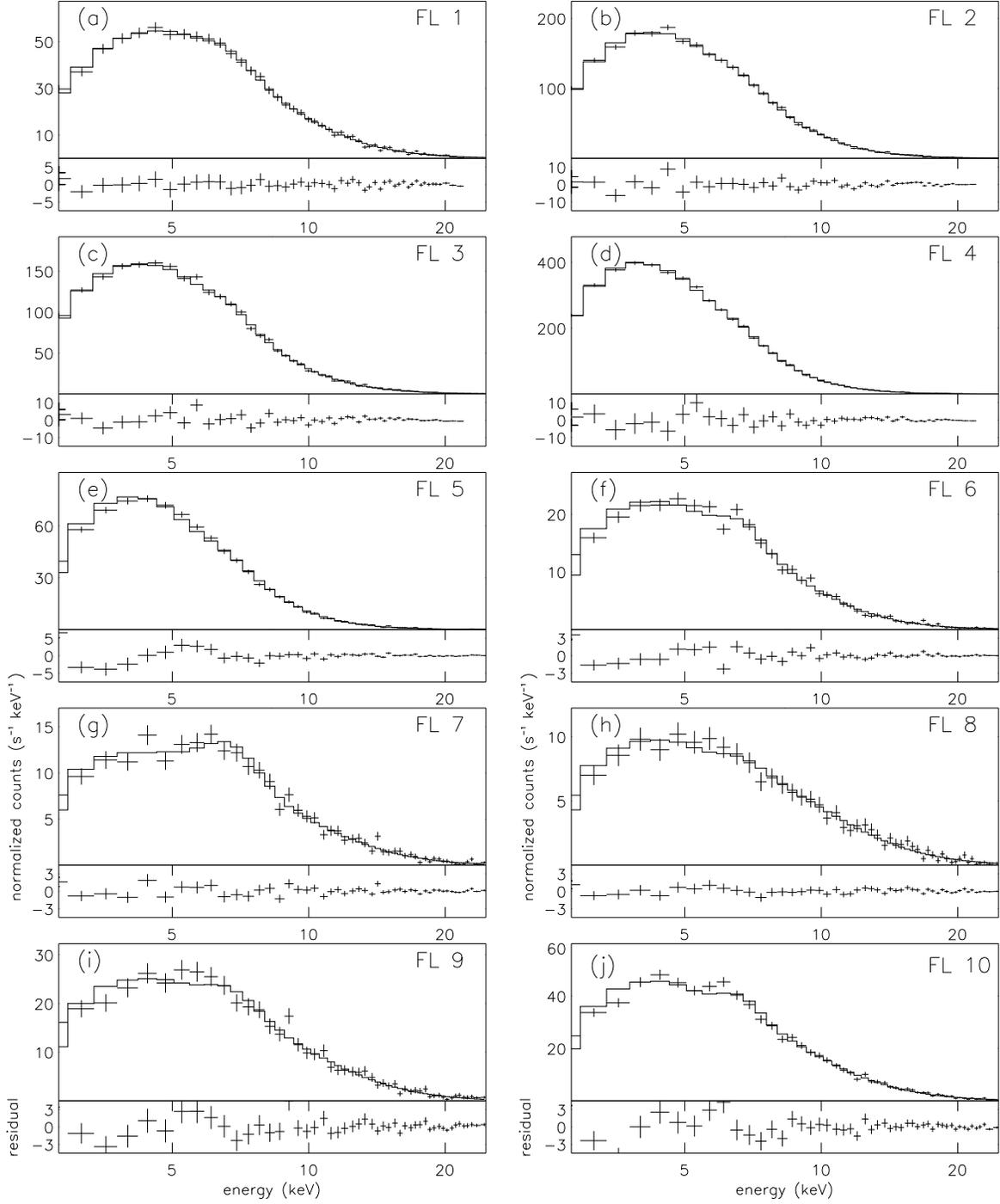}
\caption{Comparisons between the observed spectra (crosses) and the
best-fit spectra (solid lines) of the flare peaks in FL1--10.
\label{fig14}}
\end{figure}

\clearpage
\begin{figure}
\plotone{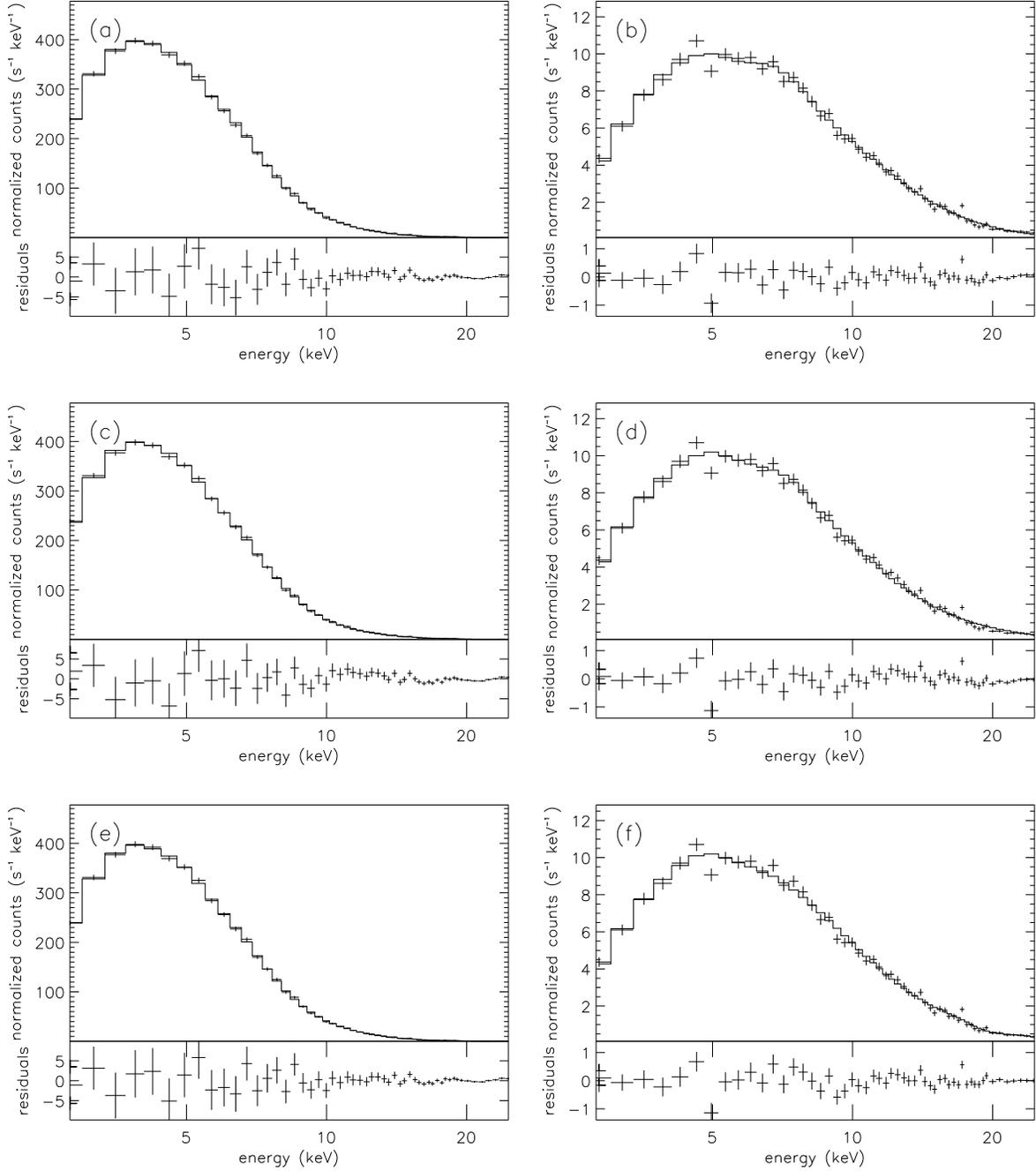}
\caption{Same as Figure 14, but for the spectra from the brightest peak in the
all flares (a, c and e), and for the spectra typical of the normal state (b, d, and f).
(a) and (b) are for model spectrum from the Eqn. (1), 
(c) and (d) are the same with (a) and (b) but without the black body component,
and (e) and (f) are for the Eqn. (2).
\label{fig15}}
\end{figure}


\begin{thebibliography}{}
\bibitem[Angelini et al. \ 1991]{aet91}
    Angelini, L., Stella, L., \& White, N. E. 1991, ApJ, 371, 332
\bibitem[Apparao \ 1991]{a91}
    Apparao, K. M. V. 1991, ApJ, 375, 701
\bibitem[Arons \ 1992]{ar92}
    Arons, J. 1992, ApJ, 388, 561 
\bibitem[Baan \ 1977]{b77}
    Bann, W. A. 1977, ApJ, 214, 245
\bibitem[Baan \ 1979]{b79}
    ------. 1979, ApJ, 227, 987
\bibitem[Basco \& Sunyaev 1976]{bs76}
    Basco, M. M., \& Sunyaev, R. A. 1976, MNRAS, 175, 395
\bibitem[Begelman \ 2001]{b01}
    Begelman, M. C. 2001, ApJ, 551, 897
\bibitem[Begelman \ 2002]{b02}
    ------. 2002, ApJ, 568, L97
\bibitem[Bildsten \& Brown \ 1997]{bb97}
    Bildsten, L., \& Brown, E. F. 1997, ApJ, 477, 897
\bibitem[Bildsten et al. \ 1997]{bet97}
    Bildsten, L., et al. 1997, ApJS, 113, 367
\bibitem[Boroson et al. \ 1999]{bet99}
    Boroson, B., Kllman, T., McCray, R., Vrtilek, S. D., \& Raymond, J. 1999, ApJ, 519, 191
\bibitem[Bradt et al. \ 1993]{bet93}
    Bradt, H. V., Rothschild, R. E., \& Swank, J. H. 1993, A\&AS, 97, 355
\bibitem[Brinerd \& Meszaros \ 1991]{bm91}
    Brainerd, J. J., \& M$\acute e$sz$\acute a$ros, P. 1991, ApJ, 369, 179
\bibitem[Brown \& Bildsten \ 1998]{bb98}
    Brown, E. F., \& Bildsten, L. 1998, ApJ, 496, 915
\bibitem[Cannizzo \ 1996]{c96}
    Cannizzo, J. K. 1996, ApJ, 466, L31
\bibitem[Cannizzo \ 1997]{c97}
    ------. 1997, ApJ, 482, 178
\bibitem[Fishman et al. \ 1996]{fet96}
    Fishman, G. J., Kouveliotou, C., van Paradijs, J., Harmon, B. A., Paciesas, W. S., Briggs, M. S., Kommerrs, J. M., \& Lewin, W. H. G. 1996, IAC Circ., No. 6272
\bibitem[Gammie \ 1998]{gam98}
    Gammie, C. F. 1998, MNRAS, 297, 929
\bibitem[in 't Zand et al. \ 1998]{iet98}
    in 't Zand, J. J. M., Heise, J., Muller, J. M., Bazzano, A., Cocchi, M., Natalucci, L., \& Ubertini, P. 1998, A\&A, 331, L25
\bibitem[Jahoda et al. \ 1996]{jet96}
    Jahoda, K., Swank, J., Giles, A. B., Stark, M. J., Strohmayer, T., \& Zhang, W. 1996, Proc. SPIE, 2808, 59
\bibitem[Kelley et al. \ 1983]{ket83}
    Kelley, R. L., Jernigan, J. G., Levine, A., Petro, L. D., \& Rappaport, S. 1983, ApJ, 264, 568
\bibitem[King et al. \ 2001]{ket01}
    King, A. R., Davies, M. B., Ward, M. J., Fabbiano, G., \& Elvis, M. 2001, ApJ, 552, L109
\bibitem[Kline et al. \ 1996]{ket96a}
    Klein, R. I., Arons, J., Jernigan, G., \& Hsu, J.-L. 1996a, ApJ, 457, L85
\bibitem[Kline et al. \ 1996]{ket96b}
    Klein, R. I., Jernigan, G., Arons, J., Morgan, E. H., \& Zhang, W. 1996b, ApJ, 469, L119
\bibitem[Kommers, Chakrabarty, \& Lewin \ 1998]{kcl98}
    Kommers, J. M., Chakarbarty, D., \& Lewin, W. H. G. 1998, ApJ, 497, L33
\bibitem[Kov\'{a}cs 2000]{k00}
    Kov\'{a}cs, G. 2000, A\&A, 363, L1
\bibitem[Kouveliotou et al. \ 1996]{ket96}
    Kouveliotou, C., van Paradijs, J., Fishman, G. J., Briggs, M. S., Kommers, J. M., Harmon, B. A., Meegan, C. A., \& Lewin, W. H. G. 1996, Nature, 379, 399
\bibitem[Kudritzki \& Puls \ 2000]{kp00}
    Kudritzki, R.-P., \& Puls, J. 2000, ARAA, 38, 613
\bibitem[La Barbera et al. \ 2001]{bet01}
    La Barbera, A., Burderi, L., Di Salvo, T., Iaria, R., \& Robba, N. 2001, ApJ, 553, 375
\bibitem[Leahy et al. \ 1983]{let83}
    Leahy, D. A., Darbro, W., Elsner, R. F., Weisskopf, M. C., Kahn, S., Sutherland, P. G., \& Grindlay, J. E. 1983, ApJ, 266, 160
\bibitem[Levine et al. \ 1991]{let91}
    Levine, A., Rappaport, S., Putney, A., Corbet, R., \& Nagase, F. 1991, ApJ, 381, 101
\bibitem[Levine et al. \ 2000]{lrz00}
    Levine, A. M., Rappaport, S. A., \& Zojcheski, G. 2000, ApJ, 541, 194 
\bibitem[Lewin, Vacca, \& Basinska \ 1984]{lvb84}
    Lewin, W. H. G., Vacca, W. D., \& Basinska, E. M. 1983, ApJ, 277, L57
\bibitem[Lewin, van Paradijs, \& Taam \ 1995]{lvt95}
    Lewin, W. H. G., van Paradijs, J., \& Taam, R. E. 1995, in X-Ray Binaries, ed. W. H. G. Lewin, J. van Paradijs, \& E. P. J. van den Heuvel (Cambridge: Cambridge Univ. Press), 175
\bibitem[Lewin et al. \ 1996]{let96}
    Lewin, W. H. G., Rutledge, R. E., Kommers, J. M., van Paradijs, J., \& Kouveliotou, C. 1996, ApJ, 462, L39
\bibitem[Li \& van den Heuvel \ 1997]{lvdh97}
    Li., X. -D., \& van den Heuvel, E. P. J. 1997, A\&A, 321, L25
\bibitem[McClintock et al. \ 1980]{met80}
    McClintock, J. E., Canizares, C. R., Li, F. K., \& Grindlay, J. E. 1980, ApJ, 235, L81
\bibitem[Moon \& Eikenberry \ 2000]{me01}
    Moon, D.-S., \& Eikenberry, S. S. 2001a, ApJ, 549, L225
\bibitem[Moon \& Eikenberry \ 2000]{me01b}
    ------. 2001b, ApJ, 552, L135 
\bibitem[Moon et al. \ 2000]{met02}
    Moon, D.-S., Eikenberry, S. S., \& Wasserman, I. M. 2002, ApJL, in press (astro-ph/0209414)
\bibitem[Naranan et al. \ 1985]{net85}
    Naranan, S. et al., 1985, ApJ, 290, 487
\bibitem[Paczy$\rm \acute n$ski \ 1992]{p92}
    Paczy$\rm \acute n$ski, B. 1992, AcA, 42, 145
\bibitem[Parmer et al. \ 1989]{pet89}
    Parmar, A. N., White, N. E., Stella, L., Izzo, C., \& Ferri, P. 1989, ApJ, 338, 359
\bibitem[Parmer et al. \ 1989]{pws89}
    Parmar, A. N., White, N. E., \& Stella, L. 1989, ApJ, 338, 373
\bibitem[Sanduleak \& Philip \ 1977]{sp77}
    Sanduleak, N., \& Philip, A. G. D. 1977, IAU Circ. 3023
\bibitem[Shakura \& Sunyaev \ 1973]{ss73}
    Shakura, N. I., \& Sunyaev, R. A. 1973, A\&A, 24, 337
\bibitem[Shapiro \& Teukolsky \ 1983]{st83}
    Shapiro, S. L., \& Teukolsky, S. A. 1983 Black Holes, White Dwarfs, and Neutron Stars: The Physics of Compact Objects
    (USA: John Wiley \& Sons), 451
\bibitem[Shaviv \ 1998]{sha98}
    Shaviv, J. J. 1998, ApJ, 494, L193
\bibitem[Shaviv \ 2000]{sha00}
    ------. 2000, ApJ, 532, L137
\bibitem[Taam et al. \ 1988]{tet88}
    Taam, R. E., Fryxell, B. A., \& Brown, D. A. 1988, ApJ, 331, L117
\bibitem[van der Klis \ 1995]{vdk95}
    van der Klis, M. 1995, in X-Ray Binaries, ed. W. H. G. Lewin, J. van Paradijs, \& 
    E. P. J. van den Heuvel (Cambridge: Cambridge Univ. Press), 252
\bibitem[van Paradijs \ 1995]{vp95}
    van Paradijs, J. 1995, in X-Ray Binaries, ed. W. H. G. Lewin, J. van Paradijs, \& 
    E. P. J. van den Heuvel (Cambridge: Cambridge Univ. Press), 536
\bibitem[Vogt \& Penrod \ 1983]{vp83}
    Vogt, S. S., \& Penrod. G. D. 1983, ApJ, 275, 661
\bibitem[Vrtilek et al. 1997]{vet97}
    Vrtilek, S. D., Boroson, B., Cheng, F. H., McCray, R., \& Nagase, F. 1997, ApJ, 490, 377
\bibitem[Woo et al. \ 1995]{wcl95}
    Woo, J. W., Clark, G. W., \& Levine, A. M. 1995, ApJ, 449, 880
\bibitem[Woo et al. \ 1996]{wet96}
    Woo, J. W., Clark, G. W., Levine, A. M., Corbet, R. H. D., \& Nagase, F. 1996, ApJ, 467, 811
\bibitem[White, Nagase, \& Parmar \ 1995]{wnp95}
    White, N. E., Nagase, F., \& Parmar, A. N. 1995, in X-Ray Binaries, ed. W. H. G. Lewin, 
J. van Paradijs, \& E. P. J. van den Heuvel (Cambridge: Cambridge Univ. Press), 1
\bibitem[White, Swank, \& Holt \ 1983]{wsh83}
    White, N. E., Swank, J. H., \& Holt, S. S. 1983, ApJ, 270, 711
\end{thebibliography}
\end{document}